\documentclass[useAMS,usenatbib,usegraphicx]{mn2e}

\newcommand{\SSS}{\scriptscriptstyle}

\title[Galaxy LMXBs]{Populating the Galaxy with Low-mass X-ray Binaries}
\author[P.D. Kiel and J.R. Hurley]{Paul D. Kiel$^{1,2}$\thanks{E-mail:
pkiel@astro.swin.edu.au (PDK)} and 
Jarrod R. Hurley$^{2}$\\
$^{1}$Centre for Astrophysics and Supercomputing, Swinburne University of Technology, Hawthorn, Victoria, 3122, Australia\\
$^{2}$Department of Mathematics, PO Box 28M, Monash University, Victoria 3800, Australia}

\begin{document}

\date{Accepted xxx. Received xxx; in original form xxx}

\pagerange{\pageref{firstpage}--\pageref{lastpage}} \pubyear{2005}

\maketitle

\label{firstpage}

\begin{abstract}
We perform binary population synthesis calculations to investigate the incidence 
of low-mass X-ray binaries and their birth rate in the Galaxy. 
We use a binary evolution algorithm that models all the relevant processes 
including tidal circularization and synchronization. 
Parameters in the evolution algorithm that are uncertain and may affect X-ray binary 
formation are allowed to vary during the investigation. 
We agree with previous studies that under standard assumptions of binary evolution the 
formation rate and number of black-hole low-mass X-ray binaries predicted by the model 
are more than an order of magnitude less than what is indicated by observations. 
We find that the common-envelope process cannot be manipulated to produce 
significant numbers of black-hole low-mass X-ray binaries. 
However, by simply reducing the mass-loss rate from helium stars adopted in the standard model, 
to a rate that agrees with the latest data, we produce a good match to the observations. 
Including low-mass X-ray binaries that evolve from intermediate-mass systems 
also leads to favourable results. 
We stress that constraints on the X-ray binary population provided 
by observations are used here merely as a guide as surveys suffer from 
incompleteness and much uncertainty is involved in the interpretation 
of results. 
\end{abstract}

\begin{keywords}
binaries: close -- stars: evolution -- stars: low-mass -- stars: neutron -- 
Galaxy: stellar content -- X-rays: binaries
\end{keywords}

\section{Introduction}
\label{s:intro}

X-ray binaries (XBs) are close binary systems that typically have luminosities in excess 
of $10^{35} \, {\rm ergs} \, {\rm s}^{-1}$ (Cowley et al. 1987) with radiation peaked 
strongly in the X-ray region. 
They are composed of a compact star -- black hole (BH) or neutron star (NS) -- 
accreting matter from a non-compact companion. 
X-ray binary systems are traditionally divided into sub classes according to the mass of the 
donor star, $M_{\rm d}$. 
High-mass X-ray binaries (HMXBs) have $M_{\rm d} > 10 M_\odot$ while 
low-mass X-ray binaries (LMXBs) have $M_{\rm d} < 2 M_\odot$ and 
intermediate-mass X-ray binaries (IMXBs) complete the classification picture. 
Observationally LMXBs are distinguished from HMXBs, because in comparison their spectra 
are devoid of normal stellar absorption features (Ritter \& Kolb 2003). 
This is thought to be the result of the donor star being overwhelmed by an accretion disk 
and therefore information from the star is lost (Ritter \& Kolb 2003). 

Detailed X-ray surveys have discovered 150 known LMXBs in our Galaxy 
(Liu, van Paradijs \& van den Heuvel 2001). 
For a subset of these, dynamical mass estimates are available and reveal that 
at least 18 are BH-LMXBs (Orosz et al. 2004). 
Empirical estimates based on these surveys and taking into account selection effects 
suggest that the total number of BH-LMXBs in our Galaxy is in the order of $1\,700$ 
(Romani 1998; Kalogera 1999). 
The observations also indicate that the formation rates of NS-LMXBs 
and BH-LMXBs are similar (Cowley 1992; Portegies Zwart, Verbunt \& Ergma 1997) 
and of the order of $10^{-7} \, {\rm yr}^{-1}$ (Kalogera 1999; Pfahl, 
Rappaport \& Podsiadlowski 2003). 
However, as previously noted by Kalogera \& Webbink (1998) the current state 
of the observational data provides only limited constraints as the data 
suffers from incompleteness and selection effects which are poorly 
understood. 
Cheif among these is the need to compensate for the potential 
transient behaviour of the LMXB sources which can produce 
variability on timescales as low as days (Chen, Scrader \& Livio 1997). 
Until we have a better understanding of the selection effects the 
empirical estimates can only be used as a general guide to the properties of 
the Galactic LMXB population. 
Even so, these estimates have been used as constraints on models of 
LMXB formation in the past. 
To date population synthesis studies have found it difficult to reproduce 
the suggested numbers and the formation rate of BH-LMXBs, especially as 
standard forms of the initial mass function make BH progenitors much rarer 
than NS progenitors (e.g. Portegies Zwart, Verbunt \& Ergma 1997; 
Podsiadlowski, Rappaport \& Han 2003). 
An associated problem is that the observed Galactic birthrate of binary millisecond 
pulsars (BMPs) is about $10^{-5} \, {\rm yr}^{-1}$ (Lorimer 1995) 
-- two orders of magnitude greater than the inferred birthrate of LMXBs which are 
thought to create BMPs (Kulkarni \& Narayan 1988). 
Recently Pfahl, Rappaport \& Podsiadlowski (2003) have shown that this discrepancy 
can be resolved if a large proportion of LMXBs are descendants of IMXBs and it is 
assumed that X-ray active lifetimes are reduced by X-ray irradiation of the donor star. 

The general formation scenario for an LMXB starts with a high-mass ($M > 10 M_\odot$) 
primary star and a low-mass ($M < 2 M_\odot$) secondary star in an orbit that is wide 
enough to allow the more massive star to evolve onto the giant branch before filling its 
Roche-lobe. 
Owing to the extreme mass ratio the mass transfer occurs 
on a dynamical timescale and is highly unstable, so that 
a common-envelope (CE) results. 
Two paths from here can be taken; either the two stars do 
not have enough orbital energy to drive off the CE and 
coalesce, therefore not producing an LMXB, or, the helium core 
of the primary and the secondary star survive the spiral-in 
phase with a smaller final separation than initially. 
The helium star then evolves to become either a NS or BH, 
depending on the initial mass of the primary. 
Finally the secondary star must evolve to fill its Roche-lobe 
(through its nuclear evolution and/or by angular momentum 
loss from the system) and initiate stable mass-transfer to the compact companion. 
This is thought to occur via an accretion disk and the result is a LMXB system. 
Typically the donor star is thought to be a main-sequence (MS) star but 
giant and white dwarf (WD) donors are also possible. 

Within the prescription based approach of population synthesis there 
are many uncertainties involved in modelling LMXB formation, not the least 
being the ad-hoc description of CE evolution (Paczynski 1976). 
Under standard assumptions for the efficiency of the energy exchange between the 
orbit and the envelope during the CE spiral-in phase, Portegies 
Zwart, Verbunt \& Ergma (1997) found that the calculated BH-LMXB 
birthrate was only 1 per cent of the NS-LMXB rate. 
Kalogera (1999) showed that abnormally high values of the efficiency parameter, 
$\alpha_{\rm\SSS CE}$, are required to gain any agreement with the observationally 
inferred BH-LMXB birthrate. 
More recently Podsiadlowki, Rappaport \& Han (2003) included a 
more accurate description of the CE process by considering 
the detailed structure of the giant star in calculating 
the envelope binding energy. 
However, BH-LMXB formation remained problematic.
In this, and the related study of Pfahl, Rappaport \& Podsiadlowski (2003) 
for NS-LMXBs, the X-ray binary mass-transfer phase was modelled using a 
full stellar evolution code. 
While representing an important step forward in the modelling approach, 
the LMXB systems, within Pfahl, Rappaport \& Podsiadlowski (2003), 
were still generated using the traditional Monte Carlo population 
synthesis technique and aspects such as tidal evolution and the accretion disk 
were omitted. 

An interesting comment was made by Kalogera (1999) that mass-loss from stars 
during the helium star phase must be less than half of their initial mass if 
significant BH-MS binary formation is to occur. 
It now seems that helium star mass-loss rates derived from observations of 
Wolf-Rayet stars may be too high by a substantial amount (Pols \& Dewi 2002). 
Reducing these rates will lead to increased BH production (at the expense 
of NSs) and will also mean that the post-CE binaries will not expand by as 
much as the helium star evolves. 
These factors combined should help in addressing the NS/BH-LMXB 
birthrate imbalance. 
Other possibilities such as additional angular momentum loss mechanisms, 
a change in the assumptions involved in CE evolution, and flaws in the 
stellar models of massive stars, have been suggested by Podsiadlowski, 
Rappaport \& Han (2003) in response to the BH-LMXB formation problem.

With an up-to-date rapid binary evolution and population synthesis package 
the aim here is to improve our understanding of low-mass X-ray binary formation.  
The effects of various model evolutionary parameters and how modification of 
these can affect model LMXB numbers and formations rates is considered. 
In particular, for the first time we will look at how alterations 
to the helium star wind mass-loss rate affects the production of BH-LMXBs. 
Other factors such as the common-envelope model and the metallicity of the star will 
also be explored. 
An advantage of the population synthesis code is that the stellar 
evolution prescription is not limited to solar metallicity stars.
Another advantage is that tidal circularization and 
synchronization of the binary orbits are followed in detail.  
This feature has been neglected in previous studies 
(e.g. Portegies Zwart, Verbunt \& Ergma 1997) 
and is important even for circular binaries as tidal 
synchronisation will continue to affect the orbital parameters.  
Also tidal forces are necessary to remove any eccentricity 
induced in a post-SN binary prior to the onset of mass-transfer. 
Aside from the fact that tidal evolution is a natural 
consequence of close binary evolution, including a description 
of tides in population synthesis models is important as it 
affects timescales and outcomes (Hurley, Tout \& Pols 2002). 

We are also motivated to make comparisons with 
previous population synthesis efforts. 
In particular we would like to address the difficulties these 
efforts have encountered in matching the observationally derived 
birthrates and numbers of BH-LMXBs, 
and the relative formation rates of NS and BH-LMXBs.  
Here we must be particularly careful in the extent to which  
we rely on the observational results. 
Romani (1998) has suggested a plausible range of 1200 to 2400 
for the actual number of BH-LMXBs in the Galaxy. 
We would like to treat this as no better than an order of magnitude 
estimate, in that it it seems much more likely that the Galactic 
population of BH-LMXBs numbers in the thousands rather than in 
the range of tens to hundreds. 
It is this latter range into which the results of previous 
theoretical studies have fallen. 

In Section~2 we describe the binary evolution algorithm and discuss uncertain aspects 
of stellar and binary evolution that affect the production of LMXBs. 
Then in Section~3 we take a semi-analytical approach to make initial constraints on 
the parameter space for LMXB formation and how variation of key evolutionary parameters  
affects this. 
Section 4 involves the population synthesis model and the calculations of the 
formation rates and numbers for NS and BH-LMXBs. 
The results are discussed in Section~5.

\section{Binary evolution model and uncertainties}
\label{s:bmodel}

\subsection{Rapid binary evolution code}

To follow the evolution of a binary system we use the Binary Star Evolution (BSE) code 
presented by Hurley, Tout \& Pols (2002). 
This package is prescription based, in that the various aspects of the evolution that can be 
envisaged are dealt with according to the best model or belief for the outcome of that 
particular process available at the time. 
This approach has obvious shortcomings compared to following the evolution with a 
detailed code (Nelson \& Eggleton 2001, for example) but has the advantage that it 
is rapid and robust and thus ideally suited to synthesizing large binary populations. 
The BSE algorithm models stellar evolution by incorporating the single star evolution (SSE) 
package (Hurley, Pols \& Tout 2000) which is a series of analytic formulae fitted to the output 
of detailed stellar models. 
For any given mass, metallicity and age, the SSE algorithm can provide quantities such as 
luminosity, radius and core mass for a star. 
This can be calculated for all evolutionary phases from the zero-age main sequence (ZAMS) 
through to and including stellar remnants. 
The SSE algorithm also includes a prescription for mass-loss in a stellar wind and follows 
the spin evolution of the star. 

The basic picture of the BSE algorithm involves two stars evolved forward in time according 
to the SSE prescription. 
In addition the algorithm needs to check whether the stars interact with one another and 
account for all possible processes which may affect the orbital parameters. 
Tidal interactions between the stars and the resulting change in orbital angular momentum and 
eccentricity are modelled in BSE for convective, radiative and degenerate damping 
mechanisms. 
Other processes which can cause angular momentum loss, such as gravitational radiation 
in close systems and magnetic braking of stars with appreciable convective envelopes, 
are also included. 
At each timestep, after the stellar parameters have been updated, all possible changes to the 
orbital angular momentum are summed, including changes owing to mass variations, and 
the orbital parameters are updated. 
The package then checks for the onset of Roche-lobe overflow (RLOF) and repeats the 
process for the next timestep if this is not the case. 

If RLOF does occur the mass transfer timescale is found (dynamical, thermal or nuclear) and, 
depending on this and other aspects such as the evolution phase of the two stars, 
the response of the system is calculated. 
If mass-transfer occurs on a dynamical timescale then depending on the mass-ratio of the 
stars the response of the system may be to form a CE which is treated as an instantaneous 
event and the outcome will be either a merger of the stars or a detached binary. 
As a result the RLOF phase is exited and the stars (or star) are evolved as described above. 
If mass-transfer is steady then RLOF is an iterative process with the primary losing an 
amount of mass $\Delta m_{1}$ at each timestep and the secondary accepting a fraction 
$\Delta m_2$ of the matter as calculated by BSE. 
To determine whether material transferred during RLOF forms an accretion disk 
around the companion star, or instead hits the companion in a direct stream, 
BSE calculates a minimum radial distance $r_{\rm min}$, of the mass stream from 
the companion (Ulrich \& Burger 1976). 
If $r_{\rm min}$ is less than the radius of the accreting star then an accretion disk is not 
formed, otherwise it is assumed that the material falls onto the companion star from the 
inner edge of a Keplerian disk. 
Total angular momentum is conserved in this model. 
Orbital changes owing to tides, gravitational radiation and magnetic braking are 
followed during RLOF and the stellar parameters, such as the stellar radii, continue to 
be updated. 
If at the end of the timestep the primary radius is still larger than its Roche-lobe then the 
RLOF process is repeated, otherwise the binary returns to being treated as a 
detached system. 
The interested reader can find full details of the BSE algorithm in Hurley, Tout \& Pols (2002).

\subsection{Model uncertanties}

We next outline some of the uncertainties in the stellar and binary evolution model 
that impact upon the formation of LMXBs. 

\subsubsection{The NS-BH mass boundary}

In terms of the initial mass of a star it is generally accepted 
that stars of $9 M_\odot$ or less will evolve to become WD remnants. 
For these stars the core remains degenerate on the asymptotic giant branch (AGB) 
and burning does not proceed past carbon. 
Above about $9 M_\odot$ stars will end their nuclear-burning lives in a core-collapse 
supernova leaving either a NS or BH compact remnant. 
We note that the actual boundary mass between WD and NS/BH formation depends 
somewhat on metallicity and the particulars of the stellar evolution model 
(Pols et al. 1998). 
Stars initially more massive than about $20 - 25 M_\odot$ are assumed to evolve 
to a BH remnant (e.g. Podsiadlowski, Rappaport \& Han 2003), corresponding to an AGB core mass 
of about $7 M_\odot$ or greater (Hurley, Pols \& Tout 2000). 

Similar to WDs, neutron stars increase in density as their mass increases in order to generate 
the higher pressure needed to counteract the effect of gravity. 
In this way there is an upper limit for the mass of a NS just as the Chandrasekhar mass limit 
exists for WDs. 
However, for NSs the upper limit is not as clearly determined as it depends upon the strong force 
of neutron-neutron interactions, and this is not fully understood. 
The limit also changes if the NS is spinning  because a star supported by both pressure and 
rotation will be less dense and will have smaller gravity than the same star not rotating. 
As a result studies place the limiting NS mass in a relatively wide range between 
1.5 and $6 M_\odot$ (Srinivasan 2002). 

Another uncertainty in determining the maximum mass of a neutron star and thus the 
boundary mass between NS and BH formation is the issue of fallback during the 
core-collapse supernova. 
Models of core-collapse are somewhat uncertain (Fryer \& Kalogera 2001) but an 
approximate picture is that initially a neutron star remnant forms from the collapse 
of the compact progenitor, the collapse is halted somewhat by the nuclear 
forces between the neutrons which are formed within the collapse. 
This halting of the collapse creates a shock wave that propagates through the outer layers of 
the star igniting further nuclear reactions and pushing on the envelope of the star. 
Depending upon the detailed structure of the stellar envelope, the shock will eject the 
envelope or it will stall and allow some or all of the envelope to fall back onto the 
initial neutron star remnant. 
If material falls back onto the star, then the mass of the remnant increases and possibly 
leads to the creation of a BH. 
Studies suggest that progenitors with a ZAMS mass of $20 M_\odot$ or less do not experience 
any fall back while masses in excess of $42 M_\odot$ lead to direct collapse of a BH 
and intermediate progenitor masses form remnants via partial fallback 
(see Belczynski, Kalogera \& Bulik 2002).  

The maximum mass of a NS and the boundary between NS and BH formation are obviously 
key factors when comparing the ratio of NS-LMXB to BH-LMXB formation. 
Belczynski, Kalogera \& Bulik (2002) found that the SSE prescription for remnant masses was 
underestimating the NS and BH masses if fallback was taken into account. 
Thus we adopt their suggested method for calculating the masses and initially 
assume that $3 M_\odot$ is the maximum possible NS mass. 
This corresponds to a boundary initial mass between NS and BH formation 
of $21 M_\odot$ for solar metallicity stars (lower for lower metallicity stars which 
have greater AGB core-mass for a given initial mass). 

\subsubsection{Supernovae velocity kicks}

Except in the case of complete fallback the supernova explosion has an associated impulsive 
mass loss. 
If the supernova occurs in a binary the ejected mass will induce an eccentricity into the orbit 
and may even destroy the system completely (Pfahl, Rappaport \& Podsiadlowski 2003). 
There is also evidence that the supernova remnant receives a velocity kick owing to an 
asymmetry in the explosion. 
This comes, for example, from observations of pulsars which have mean velocities of 
$100 - 200 \, {\rm km} \, {\rm s}^{-1}$, far in excess of the mean for normal stars 
(Fryer \& Kalogera 2001). 
The final state of a binary system involving a supernova kick depends on the orbital 
parameters at the precise moment the supernova event occurs as well as the 
magnitude and direction of the kick velocity and the amount of mass lost 
(Kalogera 1996).  
The introduction of a kick generally increases the likelihood that a binary 
will be disrupted, although Pfahl, Rappaport \& Podsiadlowski (2003) make a pertinent point that 
some systems require a kick of appropriate magnitude and direction if they are to 
survive the supernova at all. 
There is uncertainty as to whether or not BHs receive a velocity kick 
(Podsiadlowski, Rappaport, Han 2003), especially in the case of 
complete fallback. 
Observations of systems such as the BH X-ray transient GRO J1655-40 
(Willems et al. 2005) which appears to have a large space velocity would 
seem to indicate that they do although a possible explanation is that 
the system was originally a NS binary which received a kick and 
subsequent mass transfer caused the NS to evolve to become a BH. 

Unless otherwise stated we will assume that neither NSs or BHs receive a velocity 
kick at birth. 
This is in line with previous population synthesis studies 
(Portegies Zwart, Verbunt \& Ergma 1997;  
Podsiadlowski, Rappaport \& Han 2003) and recognizes that we do not fully understand the origin 
of the kicks or the conditions under which they develop (Podsiadlowski, Rappaport \& Han 2003 
and references within). 
Also, allowing velocity kicks at birth for neutron stars and black holes 
introduces a randomness to the population synthesis process (Kalogera \& Webbink 1998)  
that may blur comparisons between distinct models. 
This is not to say that we don't recognise that SN velocity kicks are a 
physical requirement, especially in the case of neutron star production. 
In models where we do include the effect of SN velocity kicks we take the kick velocity 
$v_{k}$ from a Maxwellian distribution, 
\begin{equation}
P \left( v_{k} \right) = \sqrt{\frac{2}{\pi}} \frac{v_{k}^{2}}{\sigma_{k}^{3}} \exp ^{ -\frac{v_{k}^{2}}{2\sigma_{k}^{2}}}
\end{equation}
with, $\sigma_{k} = 190 \, {\rm km} \, {\rm s}^{-1}$ as suggested by Hansen \& Phinney (1997). 

\subsubsection{Common envelope evolution}
\label{s:comenv}

The existence of cataclysmic variables with orbital periods of the order of days 
but where the progenitor of the WD must have had a radius in excess of the 
observed size of the binary at some point in the past proved puzzling for many years. 
The solution, suggested by Paczynski (1976), involves 
the partial spiral-in of the low mass secondary towards the core of the primary star, 
both of which are held within the primary stars envelope, i.e. common-envelope evolution. 
If the spiral-in process halts before coalescence because the envelope is completely driven off, 
then a close binary system can result. 
Although we have no direct evidence of such a situation, 
something like it must be at work in order to produce close binaries containing remnant stars. 
Modelling of the process is necessarily simplistic and rather ad-hoc and some variation in the 
model details exists. 

Consider a primary star of mass $M$ which overfills its Roche-lobe on a dynamical 
timescale and results in the envelope of the primary surrounding the secondary 
star of mass $m$ and the primary core of mass $M_c$. 
The secondary star and the primary core will spiral-in towards each other and transfer 
orbital energy to the envelope with an efficiency measured by the parameter 
$\alpha_{\rm\SSS CE} = E_{\rm bind} / \Delta E_{\rm orb}$. 
This is the ratio between the initial binding energy of the envelope and the change in orbital 
energy between the initial and final configurations of the CE phase. 
If the separation of the stars at the end of the CE, $a_f$, 
is such that either the secondary star or the remnant of the primary 
fill their Roche-lobes then the objects will have coalesced during the process. 
In the BSE package the binding energy is calculated as that of the giant envelope 
immediately prior to the onset of the CE, 
\begin{equation}
E_{bind} = - \frac{G M (M - M_{c})}{\lambda R} \, 
\end{equation} 
where $\lambda$ depends on the structure of the primary star. 
The initial orbital energy is that of the secondary star and the primary core 
separated by $a_i$ (the orbital separation of the binary at the onset of CE), i.e. 
\begin{equation}
E_{orb,i} = - \frac{1}{2} \frac{G M_{c} m}{a_i} \, , 
\end{equation}
and the final orbital energy is 
\begin{equation}
E_{orb,f} = - \frac{1}{2} \frac{G M_{c} m}{a_f} \, . 
\end{equation}
Alternatively the initial configuration can take the binding energy as being between the 
envelope mass and the combined mass of the primary core and the secondary star 
(Iben \& Livio 1993; Yungelson et al. 1994) which gives  
\begin{equation}
E_{bind} = - \frac{G (M_c + m) (M - M_{c})}{a_i} \, . 
\end{equation} 
Another formulation (Webbink 1984; de Kool 1990; Podsiadlowski, Rappaport \& Han 2003) 
calculates the binding energy in the same way as the BSE method but takes the 
initial orbital energy to be 
\begin{equation}
E_{orb,i} = - \frac{1}{2} \frac{G M m}{a_i} \, . 
\end{equation} 
We refer to this as the PRH CE scheme. 

In general we use the BSE CE but the effect of the alternative schemes will 
also be considered, as will the effect of varying $\alpha_{\rm\SSS CE}$. 
The parameter $\lambda$ has generally been taken as a constant of value 0.5 
in previous studies (e.g. Hurley, Tout \& Pols 2002) but is has been pointed out that this may 
underestimate how tightly bound the envelopes of massive stars are (Dewi \& Tauris 2001). 
Podsiadlowski, Rappaport \& Han (2003) have looked at detailed models of massive stars 
and found that $\lambda$ can be as low as 0.01 for stars with $M > 25 M_\odot$. 
Since the publication of the BSE algorithm, which assumed $\lambda = 0.5$ for all 
stars, an algorithm that computes $\lambda$ based on the results of detailed 
stellar models (Pols et al. 1998) has been added to the BSE package (Pols, in preparation). 
This allows $\lambda$ to vary from star to star and as a star evolves. 
Values of $\lambda$ for massive stars are in agreement with Podsiadlowski, Rappaport \& Han (2003) 
and we will look at how different assumptions regarding 
$\lambda$ affect BH-LMXB production. 

\subsubsection{Helium star wind strength}

After the NS or BH progenitor has filled its Roche-lobe as a giant and had its envelope 
stripped during CE evolution, the exposed core will evolve for some time as a 
naked helium (nHe) star before collapsing to become a compact remnant. 
Mass loss in a stellar wind from the nHe star can be crucial in determining if the 
final remnant is a NS or BH. 
Also, if the mass is lost from the binary system then a substantial increase in 
orbital separation occurs and hinders the likelihood of the binary entering a 
LMXB phase later in its evolution. 
In the SSE (and thus BSE) package mass loss from helium stars is modelled 
using the observationally derived rate 
\begin{equation}
\dot{M}_{nHe} = 10^{-13} L^{1.5} \, M_\odot \, {\rm yr}^{-1} 
\label{e:hewind}
\end{equation} 
(Hamann, Koesterke \& Wessolowski 1995) where $L$ is the stellar luminosity. 
Improved wind models have lead Hamann \& Koesterke (1998) to suggest that 
this may be an overestimate of the actual rate by a factor of 3-5 and in fact 
Nugis \& Lamers (2000) have presented a rate which is up to an order of 
magnitude smaller for typical nHe star masses. 
In light of this uncertainty we have introduced a parameter $\mu_{\rm nhe}$ into 
Equation~\ref{e:hewind} which we will use to investigate the effect of reducing 
the helium star wind strength on LMXB production. 

\section{Constraining the Parameter space}
\label{s:pspace}

Before rushing headlong into a full population synthesis it is helpful to initially constrain 
the parameter space for LMXB formation. 
To do this we take a semi-analytic approach that closely follows the method outlined 
by Portegies Zwart, Verbunt \& Ergma (1997; see also Sec. 3.2 of Kalogera \& Webbink 1998). 
The idea is that for a binary of primary mass $M$ and secondary mass $m$ we can 
quickly ascertain the range of initial separations $a$ that likely lead to an LMXB 
phase for that particular system. 
This is completed for a range of primary masses using 
the standard set of evolutionary parameters.
The standard model may be repeated with the use of alternative parameter values 
so we may gauge which parameters LMXB formation is particularly sensitive to.
We will therefore know which evolutionary parameters 
to focus on when we move to the full population synthesis 
study and thus save valuable computational time. 

\subsection{Method of constraint}

First we use the rapid SSE code to evolve a set of massive primary stars with initial masses of 
$M_0$ = 10, 15, 20, 25, 30, 35, 40, 45, 50, 55 and 60 $M_\odot$. 
Quantities of interest such as the current mass $M$, radius $R$ and core mass $M_c$, are 
recorded as a function of the evolution time as well as the mass $M_{\rm rem}$ and radius, 
$R_{\rm rem}$, of the nHe star remnant that would result if the envelope of the star 
was removed at that time. 
If an LMXB is to form then the primary star must fill its Roche-lobe at some stage of its 
evolution and initiate CE evolution. 
The Roche-lobe radius, $R_{\rm\SSS L}$, is a function of the orbital separation, $a$ and can be 
calculated using 
\begin{equation}
R_{\rm\SSS L} = \frac{0.49 a}{0.6 + q^{2/3}\ln \left( 1 + q^{-1/3} \right)} = r_{\rm\SSS L} \left( q \right) a 
\label{e:roche} 
\end{equation}
(Eggleton 1983) where $q = m/M$ is the mass ratio of the binary. 
So at any evolution time we can set $R = R_{\rm\SSS L}$ and calculate the orbital 
separation required for the primary to fill its Roche-lobe at that instant. 
If the primary star has lost mass in a stellar wind prior to filling its Roche-lobe, and this 
mass is lost from the binary system, i.e. non-conservative mass loss, then the 
separation at RLOF will be greater than the initial separation, $a_0$, of the binary. 
The two separations are related by 
\begin{equation}
a_0 (M_0 + m) = a (M + m) 
\label{e:ncsemi} 
\end{equation}
where it is assumed that the secondary does not lose mass. 
We now have the information required to calculate $a_0$ as a function of time for 
a particular $M_0$ and $m$ combination, emphasizing that $a_0$ is the upper 
limit of the initial orbital separation required for the primary to fill its Roche-lobe 
at that time, or earlier. 

As an example we show the evolution of primary radius and $a_0$ in 
Figure~1 for the case of $M_0 = 20 M_\odot$ and $m = 1 M_\odot$. 
We have taken $Z = 0.02$ as the metallicity of the two stars. 
The maximum radius reached by the primary is $1507 R_\odot$ and occurs at 
$T = 9.83\,$Myr when the star is on the AGB. 
However, the maximum value of $a_0$, denoted $a_{\rm max}$, occurs earlier 
at $T = 8.87\,$Myr just after the star evolves off the Hertzsprung gap (HG). 
The radius at this point is $860 R_\odot$ and  $a_{\rm max} = 1310 R_\odot$. 
Obviously, if the initial separation of the system is greater than this value then the 
binary will remain detached for its lifetime with no possibility of an LMXB phase. 
Also the primary can not fill its Roche-lobe at a time later than that corresponding 
to $a_{\rm max}$ as it would already have done so in the past, if this had been 
possible.

The next step in the evolution pathway for becoming an LMXB is for the binary to 
survive the CE phase. 
This occurs if the final separation, $a_f$, at the end of the CE is such that neither 
the primary remnant (a nHe star) nor the secondary star fill their Roche-lobes. 
We can calculate the minimum possible $a_f$ required for CE survival by first 
setting $R_{\rm\SSS L} = R_{\rm rem}$ and $q = m/M_{\rm rem}$ in Eq.~\ref{e:roche} 
and then repeating this process for the secondary star filling its Roche-lobe. 
The maximum of the two values is then the minimum possible $a_f$ for the 
CE to leave a binary system and generally this corresponds to the secondary 
star calculation (as was found by Portegies Zwart, Verbunt \& Ergma 1997). 
A corresponding initial separation at the onset of the CE can then be calculated 
using the equations outlined in Section~\ref{s:comenv}. 
Combining these for the BSE scheme gives 
\begin{equation}
a_i = a_f \frac{M}{M_c m} \left[ m \frac{M_c}{M} + \frac{2 \left( M - M_c \right)}{\alpha_{\rm\SSS CE} \:\:
                                                        \lambda \:\:  r_{\rm\SSS L} \left( q \right)}  \right]  \, , 
\label{e:aibse} 
\end{equation}
where $M_c = M_{\rm rem}$ and $q = m/M$. 
Here we have taken $R = a_i \, r_{\rm\SSS L}(q)$ which means we are assuming that 
the star does fill its Roche-lobe, as was assumed by Portegies Zwart, Verbunt \& Ergma (1997). 
We must then take into account any primary mass loss prior to the onset of 
CE using Eq.~\ref{e:ncsemi} to calculate an initial minimum separation (which we shall 
call $a_{0m}$) corresponding to  $a_i$. 
This is the minimum possible initial separation for the binary to survive the CE 
and therefore we require $a_{0m} < a_0$ at some point in the evolution of the 
primary if a CE is to be initiated and survived. 
If we define $a_{\rm min}$ as the separation at the point in the evolution of the primary where the 
condition $a_{0m} = a_0$ is met for the first time then the range of separations 
between $a_{\rm min}$ and $a_{\rm max}$  are of interest for LMXB formation. 
Clearly, if a system has $a_{0m} > a_0$ at all times then $a_{\rm min}$ is 
undefined and there are no possible initial separations that can lead to a 
successful CE outcome. 
The evolution of $a_{0m}$ is shown in Figure~1 for the example system and we 
see that $a_{\rm min} = 1214 R_\odot$ Myr. 
We thus find 
that systems starting with $M = 20 M_\odot$, $m = 1 M_\odot$ and a separation in the 
range $1214$ -- $1310 R_\odot$ will possibly lead to LMXB formation.
This calculation was done using the BSE CE scheme, $\alpha_{\rm\SSS CE} = 1.0$ 
and $\lambda = 0.5$.  

\begin{figure}
  \includegraphics[width=84mm]{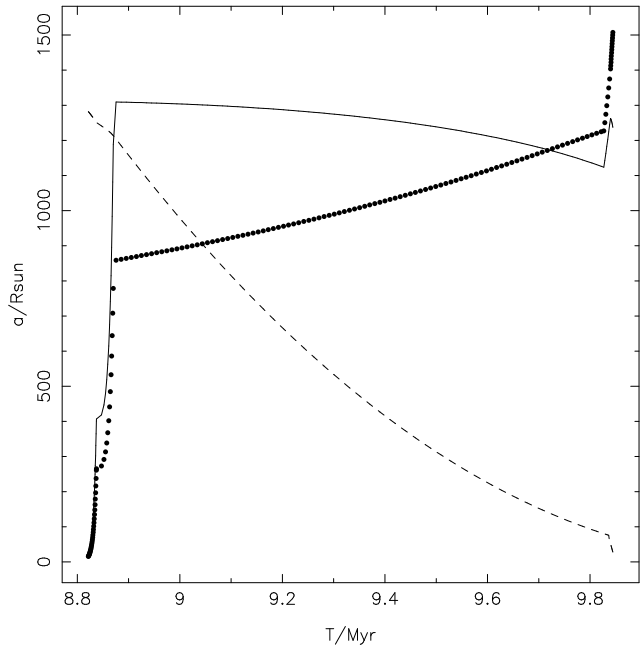}
  \caption{
  Evolution of radius (solid dots) for a $20 M_\odot$ star of solar metallicity. 
  Shown are the HG, core-helium burning and AGB phases. 
  Also plotted are the associated values of $a_0$ (solid line) and 
  $a_{0m}$ (dashed line) assuming a $1 M_\odot$ secondary (see text for details). 
   \label{f:fig1}}
\end{figure}

We note that setting $R = a_i \, r_{\rm\SSS L}(q)$ when calculating $a_i$ from $a_f$ is not formally 
correct if $a_{0m} > a_0$ because this means that the star does not 
fill its Roche-lobe for that value of $a_i$. 
In effect the binding energy of the CE has been under-estimated in the calculation 
but the bearing on the final result is not important as these cases are not of interest. 
Also, for points in the evolution where $a_{0m} < a_0$ (those that define the 
separation range) the assumption of $R = a_i \, r_{\rm\SSS L}(q)$ is valid. 

For a set of model assumptions the constraint process can be repeated for various 
primary masses in order to find the range of possible separations where, 
in later evolution, an LMXB phase may occur. 
This method is based on a fairly simplified picture of binary evolution 
and neglects aspects of the evolution such as tides which, if included, 
would be expected to shift the calculated separation range. 
However, it should not alter the existence or non-existence of a range, 
whatever the particular case may be, as tides will affect the calculation of 
$a_0$ and $a_{0m}$ in the same way. 
Processes which act on the post-CE binary are not considered in this 
semi-analytical calculation and these may conspire against LMXB 
formation even if a range of possible separations was found. 
Examples include supernovae kicks and helium star mass-loss which 
increases the orbital separation. 
What the semi-analytical process is good for is comparing results when an 
uncertain parameter in the model is varied as this gives an excellent 
indication of what values will work in favour of LMXB formation.

\subsection{Constraint results}

To compare the predicted separation ranges as a function of primary mass we will 
first define a standard model which assumes solar metallicity ($Z = 0.02$), a 
secondary mass of $1 M_\odot$, $\alpha_{\rm\SSS CE} = 1$, $\lambda = 0.5$ and the 
BSE CE scheme. 

Figure~2 shows the separation ranges of the standard model for all the 
primary masses considered. 
The separation ranges for masses less than $20 M_\odot$ are substantially larger 
than for the more massive stars. 
The $15 M_\odot$ star, for example, has $a_0$ less than $a_{0m}$ during its 
HG and core-helium burning (CHeB) phases, so no separation range is contributed here, 
however, on the AGB $a_0$ becomes less than $a_{0m}$ with $a_0$ greater 
than all previous values. 
This creates the separation range seen in Figure~2. 
In contrast the $20 M_\odot$ star has a small separation range for LMXB formation 
while in the CHeB phase but on the AGB $a_0$ is always less than previous values.  
Mass-loss truncates the AGB evolution of the $20 M_\odot$ star early in the phase 
whereas the $15 M_\odot$ star has a relatively extended AGB lifetime, and can 
evolve to large radii as a result. 
The $60 M_\odot$ star becomes a nHe star on the HG owing to substantial mass-loss 
in a wind and has $a_0$ less than $a_{0m}$ for its preceeding HG evolution. 
This is representative of stars more massive than $50 M_\odot$ which evolve too quickly 
for Roche-lobe contact to be achieved. 
Portegies Zwart, Verbunt \& Ergma (1997) did find a small separation range leading to LMXB formation 
for these stars but they used a different set of stellar models. 

\begin{figure}
  \includegraphics[width=84mm]{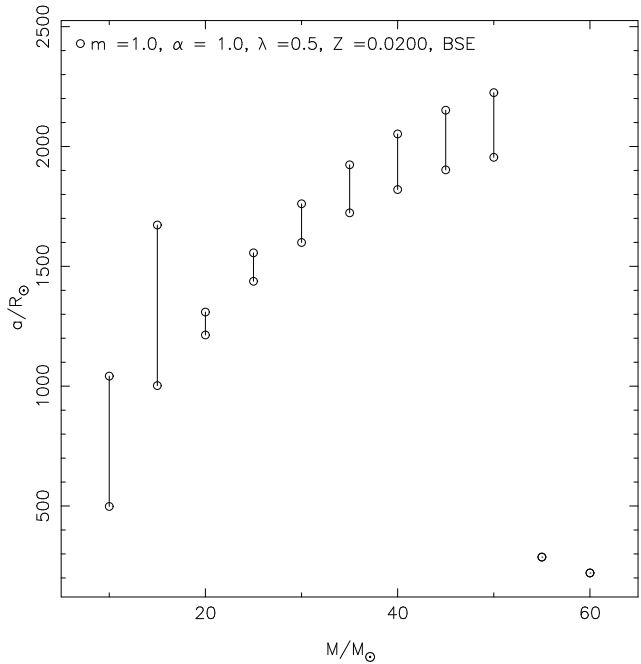}
  \caption{The separation ranges for given primary masses for the standard scenario. 
                  We note that for stars where no separation range is produced the point 
                  shown is $a_{\rm max}$ for that star. 
    \label{f:fig2}}
\end{figure}

If we switch to using the PRH scheme then 
the separation at the onset of RLOF is given by 
\begin{equation}
a_i = a_f \frac{M}{M_c m} \left[ m + \frac{2 \left( M - M_c \right)}{\alpha_{\rm\SSS CE} \:\: \lambda \:\:
                                                        r_{\rm\SSS L} \left( q \right)}  \right]  \, , 
\label{e:aiprh}  
\end{equation} 
which can be compared to Eq.~\ref{e:aibse} for the BSE CE scheme.  
This gives separation ranges that are essentially the same as those shown 
in Figure~2 for the BSE scheme, for all primary masses considered. 
Some slight differences between the two schemes naturally arise because the primary mass 
is used in calculating the initial orbital energy for the PRH CE scheme rather than the 
primary core mass that the BSE CE scheme makes use of. 
This means that for any given CE scenario the BSE scheme will lead to a greater 
value of $a_f$ than calculated using the PRH scheme, or conversely a smaller 
value of $a_i$ (for a given $a_f$). 
Basically, owing to the initial configuration of the BSE CE setup it is easier for the 
cores to drive off the envelope and avoid a merger. 
However, the choice of scheme does not make a lot of 
difference to the result. 

We next look at how the secondary mass affects the separation ranges for each star. 
In the standard model we assumed $m = 1 M_\odot$ and in Figure~3 we compare 
this result to that obtained if we instead take $0.1 M_\odot$. 
We see that only the $15 M_\odot$ star has a valid separation range when the 
secondary mass is $0.1 M_\odot$. 
Looking at Eq.~\ref{e:aibse} we see that a decrease in $m$ leads to an increase 
in the calculated $a_i$ corresponding to a particular $a_f$. 
Thus $a_{0m}$ increases with descending secondary mass. 
For example, the dashed line in Figure~1 showing how $a_{0m}$ behaves for a 
$20 M_\odot$ primary will be shifted upwards. 
Conversely, the calculated value of $a_0$ at any point in the evolution decreases 
because the mass-ratio used in Eq.~\ref{e:roche} is now smaller. 
This growth in $a_{0m}$ and the decrease in $a_0$ 
means that the initial minimum separation becomes 
less than the maximum initial separation at a later time as compared to the 
standard scenario. 
For the primary masses in which no separation range occurs the time delay 
-- which is of the order of $0.3 \,$Myr -- is significant enough for 
$a_0$ to start decreasing 
by the time that $a_{0m} < a_0$ first occurs. 
The difference for the $15 M_\odot$ primary star is its extended AGB lifetime where 
$a_0$ becomes larger than previous values while also larger than $a_{0m}$. 
The lower limit to the separation range has decreased because of the decrease in $a_0$ 
corresponding to a particular value of the primary radius. 
For the $10 M_\odot$ star the separation range has disappeared because in this 
case the decrease in $a_0$ values ensures that $a_{0m} > a_0$ at all points in the 
evolution. 

\begin{figure}
  \includegraphics[width=84mm]{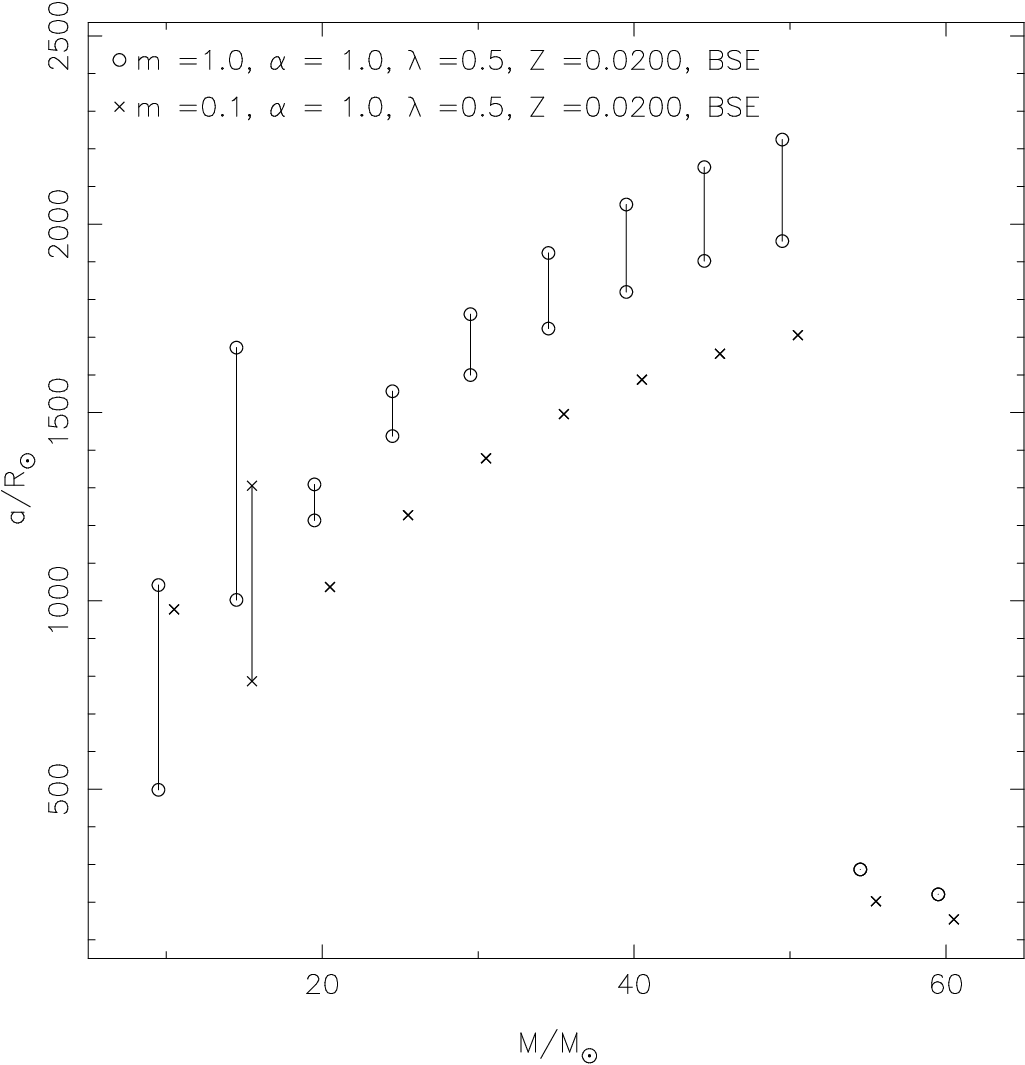}
  \caption{Comparison of separation ranges for given primary masses 
    between the standard scenario and a change to $m = 0.1 M_\odot$.
    \label{f:fig3}}
\end{figure}

Figure~4 exhibits how increasing $\alpha_{\rm\SSS CE}$ from 1.0 to 3.0 affects 
the results. 
This corresponds to a decrease in the calculated $a_{0m}$ values which leads 
to an increase in the separation ranges found for all masses. 
We now see that ranges even exist for the $55$ and $60 M_\odot$ primaries. 
On the other hand a decrease in $\lambda$ (or equivalently $\alpha_{\rm\SSS CE}$) 
means that the orbit must generate more energy to overcome the envelope 
binding energy and as a result this makes it more likely that a CE event will 
end in a merger. 
As mentioned in Section~2.2, Podsiadlowski, Rappaport \& Han (2003) show that $\lambda$ 
can be as low as 0.01 during stages of massive star evolution and in Figure~5 
we investigate how lowering $\lambda$ from 0.5 to 0.2 
affects the potential for LMXB formation. 
We see that even this conservative decrease leads to the separation ranges 
vanishing for all stars except, once again, the $15 M_\odot$ star. 

\begin{figure}
  \includegraphics[width=84mm]{fig4pk.eps}
  \caption{Comparison of separation ranges for given primary masses 
    between the standard scenario and a change to $\alpha_{\rm\SSS CE} = 3$.
    \label{f:fig4}}
\end{figure}

\begin{figure}
  \includegraphics[width=84mm]{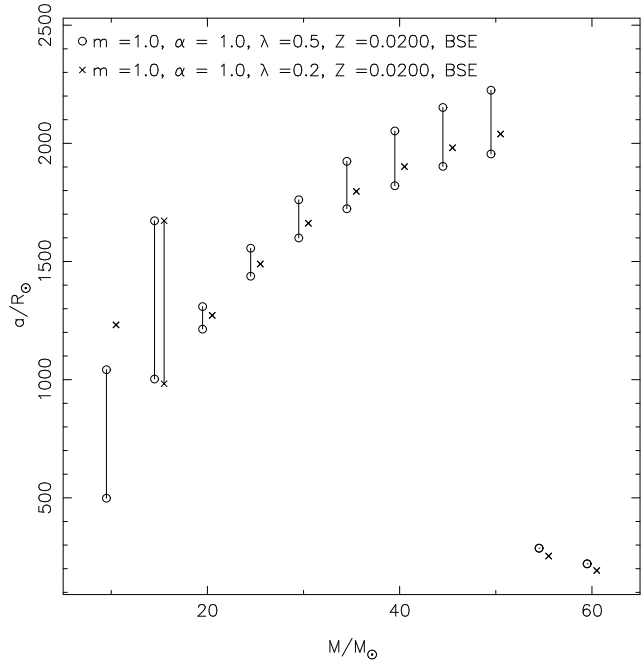}
  \caption{Comparison of separation ranges for given primary masses 
    between the standard scenario and a change to $\lambda = 0.2$.
    \label{f:fig5}}
\end{figure}

Finally we present Figure~6 which compares $Z = 0.02$ and $0.0001$. 
The major factor producing the separation range differences that we see is 
that radius evolution for massive stars changes markedly with metallicity. 
For example, a $20 M_\odot$ star with $Z = 0.0001$ has a radius of only 
$20 R_\odot$ when it reaches the end of the HG and the corresponding 
value of $a_0$ is $35 R_\odot$, much less than for the solar case, while 
$a_{0m}$ remains high. 
The low-$Z$ star does not have $a_{0m} < a_0$ until it nears the AGB, 
with a radius of about $800 R_\odot$. 
So, compared to the solar metallicity star, the separation range for LMXB formation 
occurs much later in the evolution of the star and leads to an extended range 
but at smaller separations. 
For higher mass stars differences in mass-loss rates can also become important. 
A $40 M_\odot$ solar metallicity star loses about $5 M_\odot$ of material in a wind 
while on the MS compared to only about $0.3 M_\odot$ for its low metallicity 
counterpart. 
So the value of $a_0$ calculated at similar points in the evolution will be smaller 
for the low-metallicity star, not only because its radius is smaller but also because 
the mass-ratio used in Eq.~\ref{e:roche} will be smaller. 
For the $40 M_\odot$ low-metallicity star, and higher masses, we find that $a_{0m} > a_0$ 
at all points in the evolution, and thus no range is found.

\begin{figure}
  \includegraphics[width=84mm]{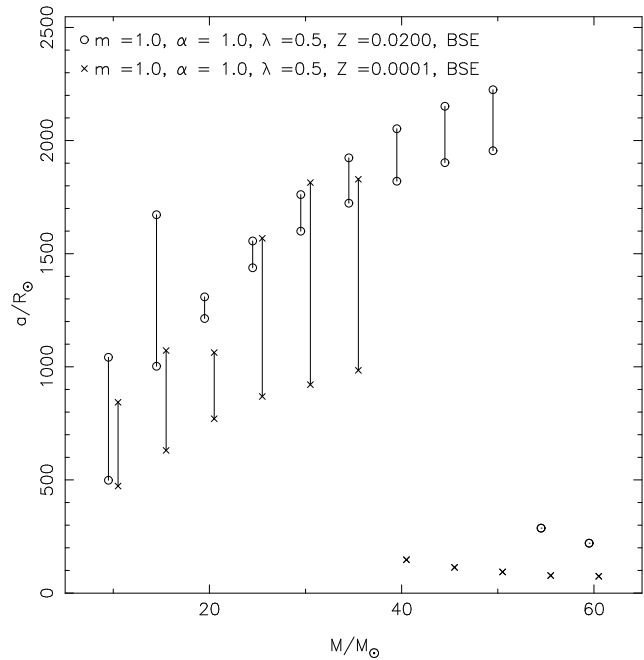}
  \caption{Comparison of separation ranges for given primary masses 
    between the standard scenario and a change to $Z = 0.0001$.
    \label{f:fig6}}
\end{figure}

As expected, the common-envelope efficiency parameter has the greatest 
potential to increase the incidence of LMXB formation. 
Our preliminary studies have shown that we should not expect significant 
LMXB formation, especially not BH-LMXB formation, from low values of 
$\lambda$ or very low-mass secondaries. 
This agrees with the findings of Podsiadlowski, Rappaport \& Han (2003). 
The effect of metallicity on the formation rate of LMXBs will also be 
interesting to study with full population synthesis.

\subsection{An evolution example} 

We next give an overview of LMXB evolution for a particular binary 
and then illustrate how the evolution changes according to some of 
the choices made for the parameters of the BSE code. 
Consider a binary system which involves a primary star with an initial mass 
of $45 M_\odot$ and a secondary star with an initial mass of $1 M_\odot$. 
The system begins with a period of 1705 days, or approximately a 
separation of $2\,150 R_\odot$, and initially has a circular orbit. 
The PRH CE scheme is used and the model parameters $\alpha_{\rm\SSS CE}$ and 
$\lambda$ are assumed to be 1.0 and 0.5 respectively. 
The metallicity is of solar abundance, $Z = 0.02$.

Initially the evolution of the primary star, relative to the secondary star, 
is very fast: the MS lifetimes are 4.5 and $10\,950\,$Myr respectively. 
While the massive primary is on the MS it loses $6.5 M_\odot$ of mass due 
to a wind and the orbital separation slowly increases as this mass 
is lost from the binary system, taking angular momentum with it. 
When the primary reaches the end of the MS the separation is $2505 R_\odot$. 
The primary then evolves quickly along the HG all the while increasing its radius. 
Only $0.01$ Myr after leaving the MS the primary's radius reaches and exceeds 
its Roche-lobe radius and mass transfer is intiated. 
The secondary star is still on the MS.  
At the onset of RLOF the primary has a mass of $38.4 M_\odot$ 
and it has just become hot enough for core helium burning to begin 
-- stars initially more massive than about $12 M_\odot$ can ignite helium while on the HG 
and effectively do not have a first giant branch (see Hurley, Pols \& Tout 2000). 
The mass-transfer proceeds on a dynamical timescale, which produces a 
runaway effect, where the 
entire envelope of the CHeB star (mostly convective) overflows its Roche-lobe and forms a CE 
around the helium core of the primary and the MS secondary. 
This is assumed to occur instantaneously. 
The helium core and the MS star spiral-in towards each other and drive off the 
envelope before getting close enough to coalesce (see section 2.2.3 for CE description).

The CE phase leaves the binary with a circular orbit and a separation 
of $6 R_\odot$. 
The primary star, now a naked helium MS star, has a mass of $13.6 M_\odot$ 
and its radius is well within its Roche-lobe radius. 
The secondary star, however, is more likely to fill its Roche-lobe now, 
as the separation has decreased immensely. 
While the secondary is still on the MS, the primary naked helium MS star 
evolves onto the helium HG at a system time of 5.2 Myr with a mass of 
$8.2 M_\odot$. 
Shortly afterwards the primary star goes SN. 
The outcome of the SN event is a $4.7 M_\odot$ BH primary. 
The mass lost during the SN is assumed to occur instantaneously and is 
not enough to disrupt the binary. 
It did, however, expand the orbit to a separation of $23 R_\odot$ and 
induced an eccentricity of 0.58. 
In this scenario a SN kick for BH formation was not assumed.
The mass of the secondary star has increased to $1.06 M_\odot$ owing 
to mass accretion from the wind of the primary while it was a 
helium star. 

Jumping ahead to a system time of $11\,741\,$Myr the secondary begins RLOF 
and the binary is an LMXB. 
The secondary is on the GB, so the system is not a typical LMXB if the 
strict classification requiring the secondary to be a MS star is 
applied. 
The system is circularised owing to tidal forces acting upon the orbit 
and at RLOF has a separation of $14 R_\odot$. 
Mass transfer occurs on a nuclear timescale and lasts for $\sim 300\,$Myr. 
The BH mass, in this time, increases to $5.4 M_\odot$, while the 
secondary mass decreases to $0.49 M_\odot$. 
During the LMXB phase the X-ray luminosity peaked at about $66\,500 L_\odot$ and 
the mass-transfer ends when the giant shrinks within its Roche-lobe radius 
with only $0.01 M_\odot$ remaining in its envelope. 
The separation at the end of the LMXB phase has increased to $109 R_\odot$ 
and stays, roughly, at this distance for the rest of the system lifetime. 
The secondary ends up as a white dwarf with a radius well within its Roche-lobe radius.

We note that if we neglect tidal forces for our example binary then the 
orbit remains eccentric when the LMXB mass-transfer phase begins 
and the separation between the stars is also greater which means 
that RLOF actually starts later at a time of $11\,909\,$Myr. 
The LMXB phase lasts for about $150\,$Myr which is a factor of two 
less than when tides are included. 

It is interesting and insightful to consider how the evolutionary 
pathway changes when $\alpha_{\rm\SSS CE}$ (or equivalently $\lambda$) 
and $\mu_{ \rm nhe}$ are varied within this example system. 
Changing $\alpha_{\rm\SSS CE}$ to 3 gives a wider post-CE binary with a separation of 
$18 R_\odot$. 
Consequently the LMXB phase does not start until a time of 12215 Myr and 
lasts for only 56 Myr. 
The final BH mass is $5.3 M_\odot$ and the mass of the secondary at the 
end of the LMXB phase is $0.36 M_\odot$. 
Decreasing the common-envelope efficiency leads to a tighter post-CE 
system and increases the chances of a merger occurring during CE. 
This is equivalent to a decrease in $\lambda$, the binding energy parameter, 
and indeed if we use $\lambda = 0.2$ in our example then the helium 
core of the primary and the secondary coalesce during CE and there is no LMXB 
phase. 
We note that if a velocity kick had been allowed at birth for the 
BH then the binary would have been destroyed at this time. 

We next evolve this example binary according to the standard parameters 
but with the helium star wind strength reduced by a factor of 5, 
i.e. $\mu_{\rm nhe} = 0.2$. 
Decreasing the wind mass loss from helium stars lessens the increase in 
the orbital separation during this phase. 
We now find that when the primary ends its helium star lifetime 
it becomes a BH of mass $11.9 M_\odot$ and the separation is only 
$6.7 R_\odot$. 
As a result the secondary fills its Roche-lobe while still on the MS 
at a time of $1\,346\,$Myr and a long-lived classical LMXB phase 
continues to a time of $15\,000\,$Myr and beyond. 
It is obvious that reducing the strength of the helium wind is 
desirable in terms of producing long-lived BH-MS-LMXBs and thus will 
aid in increasing the formation rate and numbers of the binaries. 
This will be explored in the next section.

\section{Population synthesis and Results}
\label{s:results}

With the evolutionary parameters considered in depth, a statistical 
description of LMXBs is now investigated.
This statistical approach is in the form of a full population synthesis 
and is used to evolve many ($\sim 10^6$) binaries, producing an 
estimation for the rate of birth of LMXBs and the 
numbers of these systems in the Galaxy now.

To populate the Galaxy with binaries we first produce a grid of suitable 
initial parameters (primary mass, secondary mass and orbital separation)  
and evolve each binary on the grid. 
For each binary we take into account realistic initial mass functions (IMFs) 
and period distributions to determine the likelihood of the binary 
contributing to the Galactic population. 
Then by picking out the systems that evolve to become LMXBs, and 
combining this with a star formation rate, we find the birthrate and 
expected number of LMXBs in the Galaxy. 
This is repeated for a variety of models.

\subsection{Method}

A binary system can be described by three initial parameters: primary 
mass, M, secondary mass, m, and separation, a (or period). 
A fourth parameter is the eccentricity but it has been shown previously 
that this has little affect on population synthesis results (Hurley, Tout \& Pols 2002). 
We assume circular orbits here for simplicity. 
For the primary mass we choose a minimum mass of $5 M_\odot$ and a 
maximum mass of $80 M_\odot$. 
The lower limit takes into account the requirement that the primary must form 
a NS or BH to be of interest and $5 M_\odot$ is considered to be safely below 
the minimum initial mass for NS formation (noting that we will be evolving some 
low-metallicity models). 
As suggested in Hurley, Tout \& Pols (2002) the production of stars with 
$M > 80 M_\odot$ is rare when using any reasonable IMF, hence the choice 
of upper mass limit. 
We consider secondary masses between the limits of $0.1$ and $2.0 M_\odot$. 
The upper mass limit in this case is based on the usual definition that an 
LMXB has a donor star of $2.0 M_\odot$ or less, while the lower limit 
is approximately the minimum mass of a star that will ignite hydrogen-burning 
on the MS. 
For the initial orbital separation we consider a range between $10$ to 
$3000 R_\odot$. 
This choice is guided by the results of the previous section with some 
leeway at either end. 
We did maintain a check while evolving each model to see if the upper limit 
was breached at any point in the binary evolution but even though some 
LMXB systems were produced from initial separations close to the upper limit, 
none were produced from binaries with $a_0 = 3000 R_\odot$. 
So the choice proved to be fine. 

For the primary mass we chose to have 100 grid points and to have these 
logarithmically spaced so that the grid was finer at lower masses. 
This gave step lengths of  $\Delta \log\left( M \right) = 0.012 M_\odot$. 
We chose to use 50 grid points for the secondary mass with a linear 
spacing of $\Delta m = 0.04 M_\odot$, 
For the separation we chose 600 steps with $\Delta a = 5 R_\odot$. 
This gave a total of $3\times10^6$ distinct initial parameter sets and 
each binary on the grid was evolved with the BSE algorithm to an age 
of $15\,$Gyr -- deemed to be a reasonable upper limit to the likely 
age of the Galaxy. 

To calculate birthrates we must combine our grid of initial parameters 
with realistic distribution functions in order to calculate the contribution 
of each binary to the rate. 
In selecting the form of the initial distributions we are guided by the choices 
made by Hurley, Tout \& Pols (2002) and indeed the process of 
calculating birthrates and expected numbers follows closely the process 
described in that work. 
For the distribution of primary masses we use the 
power law IMF given by Salpeter (1955),
\begin{equation}
\Phi \left( M \right) = kM^{-\alpha},
\end{equation} 
with $\alpha = 2.35$ and $k = 0.0603$. 
This is the probability that a star has mass M. 
There is some uncertainty in the slope of this distribution with 
Kroupa, Tout \& Gilmore (1993) finding that for masses larger than 1.0 $M_\odot$ 
a value of $\alpha = 2.7$ is a better fit to solar neighborhood data. 
The distribution of secondary star masses is not as well constrained and we 
simply take 
\begin{equation}
\varphi \left( m \right) = \frac{1}{M}.
\end{equation} 
as previous authors have done 
(Portegies Zwart, Verbunt \& Ergma 1997; Hurley, Tout \& Pols 2002). 
This assumes that the initial secondary and primary masses are closely related 
via a uniform distribution of the mass-ratio. 
The initial separation distribution is taken to be flat in log(a),
\begin{equation}
\Psi \left( \log \left( a \right) \right) = {\rm b}, 
\end{equation} 
where $b = 0.35993$ is a constant determined by normalization of the 
distribution between the chosen limits. 

For every system which went through an LMXB phase we determine a contribution 
\begin{equation}
\Delta r  = S \, \Phi \left( \log \left( M \right) \right) \Delta \log \left( M \right) 
                      \varphi \left( m \right) \Delta m \, \Psi \left( a \right)\Delta a 
\end{equation} 
(Hurley, Tout \& Pols 2002) to the overall birthrate. 
Here $S$ is the star formation rate which we take to be $7.6 M_\odot \, {\rm yr}^{-1}$. 
This is in rough agreement with Yungelson, Livio \& Tutukov (1997) who quote 
$8.3 M_\odot \, {\rm yr}^{-1}$ but note that if the assumed minimum mass of newborn 
stars is raised from $0.1$ to $0.3 M_\odot$ the rate drops to $4.7 M_\odot \, {\rm yr}^{-1}$. 

In this manner we calculate the rate of birth for NS-MS-LMXBs and BH-MS-LMXBs 
by summing the contributions for all binaries that passed through the relevant phase 
of evolution. 
We can also estimate the numbers of LMXBs residing in the Galaxy at an 
age $T$ using  
\begin{equation}
N  =  \sum_{j} \Delta r_{j} \times \Delta t_j \, ,
\end{equation}
where we are sum over all $j$ systems that had an LMXB phase of 
length $\Delta t_j$ that began prior to $T$. 

\subsection{Models}

First we define a standard model which we shall call Model~A. 
This assumes solar metallicity, $Z = 0.02$, a helium wind parameter of 
$\mu_{\rm nhe} = 1$ and that the maximum NS mass is $M_{\rm max,NS} = 3 M_\odot$. 
For common-envelope evolution the BSE CE scheme is used in conjunction 
with $\alpha_{\rm\SSS CE} = 3.0$, and $\lambda$ is allowed to vary according to 
the new algorithm added to the BSE package (Pols, in prep.). 
The change to $\alpha_{\rm\SSS CE} = 3$ for the standard model, compared to 
the standard scenario used in Section~3 which had $\alpha_{\rm\SSS CE} = 1$, 
is prompted by our desire to maximize the formation of BH LMXBs. 
We note that such a value is not unusual (Nelemans et al. 2000) 
and that Hurley, Tout \& Pols (2002) showed that using $\alpha_{\rm\SSS CE} = 3$ in the 
BSE CE scheme is equivalent to using $\alpha_{\rm\SSS CE} = 1$ in the 
Yungelson et al. (1994) scheme. 
A velocity kick is not given to NSs or BHs at birth in the standard 
model (as explained in Sec. 2.2).  
The particulars of this model, and all of the population synthesis models 
that we introduce, are summarized in Table~1. 

In Model~B we investigate the effect of using the PRH CE scheme. 
Models~C and D differ from Model~A by assuming $\alpha_{\rm\SSS CE} = 10$ 
and $\alpha_{\rm\SSS CE} = 1$, respectively (using the BSE CE scheme). 
For Model~E we lower the maximum NS mass to $M_{\rm max,NS} = 2 M_\odot$ 
in order to quantify how this affects BH LMXB formation. 
Then for Model~F we turn the helium wind mass-loss strength down by a factor 
of two so that $\mu_{\rm nhe} = 0.5$. 
After analysing this initial set of models it turns out that Model~F gives much 
better agreement with observations than Model~A (see below) and thus the 
choice was made to adopt $\mu_{\rm nhe} = 0.5$ as a standard feature in 
subsequent models.

In Models~G and H we investigate the effect of metallicity; Model~G has $Z = 0.001$ 
while $Z = 0.0001$ is assumed for Model~H. 
Both use $\mu_{\rm nhe} = 0.5$ so are directly comparable to Model~F. 
Although it is a naive assumption to have all stars within a galactic population born 
with the same metallicity, comparison of models of various metallicity will at 
least help in constraining the effect of metallicity upon LMXB populations. 
The effect of SN velocity kicks are the focus of Model~I where we use the kick 
distribution mentioned in Section~2.2.

\subsection{Results and Analysis}

In Table~2 we give the rates of birth for NS and BH-MS-LMXBs and
the estimated numbers of BH-MS-LMXBs currently in the Galaxy 
predicted by each of the models. 
We find that the standard scenario BH-MS-LMXB birth rate is less than 0.1\% 
of the NS-MS-LMXB rate, 
while the number of BH-MS-LMXB systems produced is only 17. 
Recalling that we are aiming for BH-MS-LMXB numbers 
of the order of a thousand then clearly Model~A is not suitable. 
It is this order of magnitude difference (or even greater) between 
predictions from observation and modelling which we try to 
address here through reasonable changes in the evolutionary parameters. 
We also aim to bring the relative birthrates of NS and BH-LMXBs 
to within an order of magnitude agreement. 
Tighter constraints than this are not within the scope of the 
current observations which we use merely as a guide when 
comparing various models. 

As expected from Section~\ref{s:pspace} and shown in Table~2, switching between 
the BSE and PRH CE schemes gives negligible difference between
the rate of production and numbers of BH-MS-LMXB systems within the Galaxy.
This is illustrated in Figures 7 a) and b), which plot the initial orbital separation and primary mass
of binaries which initiate a BH-LMXB phase in Models~A and B. 
The systems which are only produced by Model B, when compared to Model A, are generally 
not BH-MS-LMXBs but rather systems involving NSs. 
Thus the slight rise in NS-MS-LMXB rates for Model~B. 
There are some additional  BH-LMXB systems produced in Model~B but these 
do not contain a MS companion.
It must be noted that a NS may accrete enough matter to evolve into a BH 
and therefore the progenitor mass of the BH may be smaller than the 
progenitor NS/BH mass boundary.
For solar metallicity (both Models A and B) this initial mass boundary is at approximately 
$21 M_\odot$,
therefore the islands of parameter space below this mass in Figures~7 a) and b) 
are BH-LMXBs which have formed via accreting NSs. 

The analysis of Section~\ref{s:pspace} showed that a system with a 
high value of $\alpha_{\rm\SSS CE}$ 
has a greater chance of surviving the CE phase. 
As an illustration of the effect of this on LMXB production we take 
the extreme case of $\alpha_{\rm\SSS CE} = 10.0$ in Model~C. 
We see that the rates and numbers of BH-MS-LMXBs in Model~C 
are higher than those produced by Model A, as expected. 
The initial parameter space of binaries in Model~C that evolve to become BH-LMXBs is 
shown in Figure~7c) and the difference compared to Model~A is clearly visible. 
What stands out between these two models is the increase in LMXB systems 
which form with smaller separations in Model~C; there is a $\Delta 1\,500 R_\odot$
difference in minimum separation between the $36 M_\odot$ to $52 M_\odot$ 
primary mass range for the two models.
The LMXB systems created by Model C that induce the slight 
increase in BH-MS-LMXB numbers are long-lived BH-MS-LMXB systems. 
The number of LMXB systems, however, has not risen enough to justify such a 
large increase in $\alpha_{\rm\SSS CE}$. 
We stress that using the unphysically high value of $\alpha_{\rm\SSS CE} = 10.0$ 
was merely to illustrate that uncertainty in the common-envelope 
parameter cannot be utilised to sufficiently raise BH-MS-LMXB numbers. 
We do not consider this to be a valid parameter change. 
Looking at Table~2 we see that decreasing $\alpha_{\rm\SSS CE}$ to 1 (Model D) 
means that no BH-MS-LMXB systems are formed.
Again, from Section~\ref{s:pspace}, this is not surprising. 

Manipulating the boundary of NS and BH formation in Model E 
obviously has an effect on the numbers of NS and BH-LMXBs produced.
Not only is the progenitor NS/BH boundary mass decreased, 
therefore making it easier to create BHs, but the production of BHs 
from accretion onto NSs is now greater. 
The effect on the parameter space of binaries leading to a BH-LMXB phase 
is shown in Figure~7d) and can be compared to Figure~8a) to see the difference 
between Models~E and A. 
The progenitor mass NS/BH boundary for Model E is approximately $18 M_\odot$. 
Systems with primary masses lower than this are created via accretion onto NSs which 
then gain enough mass to collapse to a BH and the increase in these systems for 
Model~E simply reflects the decrease in maximum NS mass from 3.0 to $2.0 M_\odot$. 
This reduction in maximum NS mass however, does not noticeably affect the birth rate
of NS-MS-LMXBs but does increase the rate of BH-MS-LMXBs formation.
This change produces satisfying results, the BH-MS-LMXB formation rate is 
almost within an order of magnitude of the NS-MS-LMXB rate
while the number of BH-MS-LMXB systems lies within the predicted 
observational limits.

\begin{figure}
  \includegraphics[width=84mm]{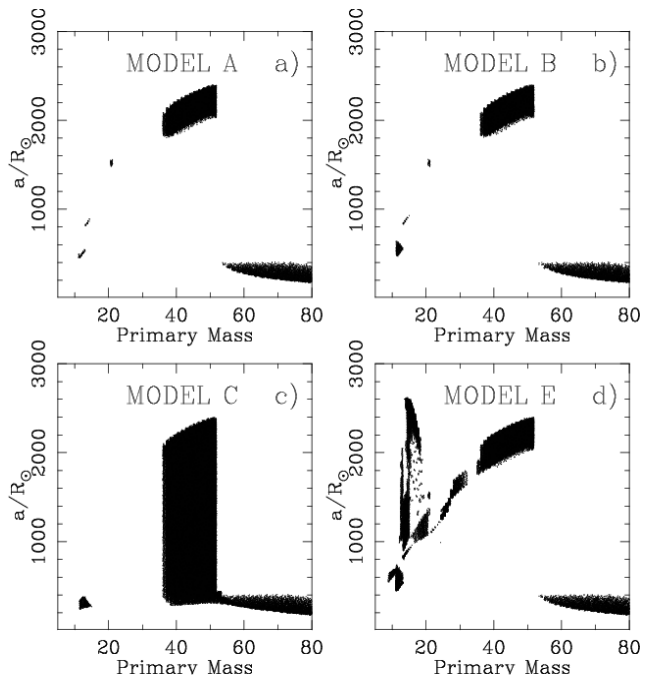}
  \caption{Parameter space of initial separation and primary mass for 
                  binaries leading to a BH-LMXB phase in Models~A, B, C and E. 
    \label{f:fig7}}
\end{figure}

Model F which reduces the helium star mass-loss rate also greatly 
enhances the BH-MS-LMXB formation rate to make it almost within an 
order of magnitude of the NS-MS-LMXB formation rate,
This is similar to the results of Model~E and is encouraging. 
In fact the BH-MS-LMXB formation rate for Model~F is a factor of 100 or more greater 
than we obtained for Model~A and is now in the range of $10^{-7} \, {\rm yr}^{-1}$ as 
indicated by observations (Kalogera 1999). 
The number of BH-MS-LMXB systems produced by Model F is also very good 
and well within the estimations from observations (see Table~\ref{t:table2}). 
Figure~8 a) depicts the region of primary mass and binary separation parameter space 
that involves BH-LMXBs for Model F. 
This region is not as extensive as in some of the previous models, yet
in comparison to these models, Model F produces a much larger proportion of BH-MS-LMXBs.
With Model~F we find that almost 90\% of the BH-LMXB systems produced involve a 
MS donor star whereas for Models~A and E only 15\% and 28\%, respectively, of 
the systems shown in Figure~7 have MS secondary stars.
Also, the BH-MS-LMXB systems created in Model~F are generaly long lived, 
therefore each system makes a greater contribution to the estimated total number. 
Consider a system with primary mass, $M = 40.8 M_\odot$, secondary 
mass, $m = 0.9 M_\odot$ and separation, $a = 1906 R_\odot$. 
In Model A this binary survives the CE phase but the post-CE binary experiences 
substantial mass-loss both from the primary helium star and during the resulting 
SN in which the primary becomes a BH. 
The separation after the SN is about $60 R_\odot$ and the low-mass MS seconday 
has no chance of filling its Roche-lobe while on the MS. 
In Model~F however the primary star loses substantially less mass while a helium star 
and also the mass of the star when it becomes a BH is greater than in Model~A. 
These effects combine to give a much closer post-SN binary with a separation of 
about $8 R_\odot$. 
Subsequently the MS secondary is able to fill its Roche-lobe and a long-lived 
BH-MS-LMXB phase results. 
We are much more comfortable with changing the helium wind parameter than we 
are with altering the maximum NS mass as for the former there is observational 
evidence suggesting the change (see Section 2.2.4). 
We now continue by adopting $\mu_{\rm nhe} = 0.5$ as standard and compare all 
subsequent models to Model~F. 

In Model~G we assume a metallicity of 0.001 for the stars and we find that 
the number of BH-MS-LMXBs has 
increased to $3\,290$ compared to $1\,870$ for 
Model~F which assumed solar metallicity.  
Figure~8b) shows the range of initial parameters leading to BH-LMXBs in 
Model~G and we find that for separations greater than about $1000 R_\odot$ 
much more of the parameter space is covered in comparison to Model~F. 
This is because with lower metallicity the massive primary stars suffer less 
mass-loss from winds so that prior to the CE event there is less increase in 
orbital separation (see Figure~\ref{f:fig6}). 
As a result binaries in Model~G will be closer than their Model~F equivalents 
and therefore more likely to interact. 
The radius evolution of the stars is also affected by a change in metallicity and 
this affects the strength of tidal forces operating in the binaries. 
There is also a further region in parameter space extending down to $39 M_\odot$ 
in primary mass for low separation values, which does not occur in Model F.
However, it is a region between $23 M_\odot$ to $40 M_\odot$ in primary mass and from 
$600 R_\odot$ to $1\,650 R_\odot$ in separation where the BH-MS-LMXB systems are created.
This region is much larger than that of Model F, which is slightly higher and narrower
in separation, resulting in a greater number of  long-lived MS-LMXB systems in Model G.

\begin{figure}
  \includegraphics[width=84mm]{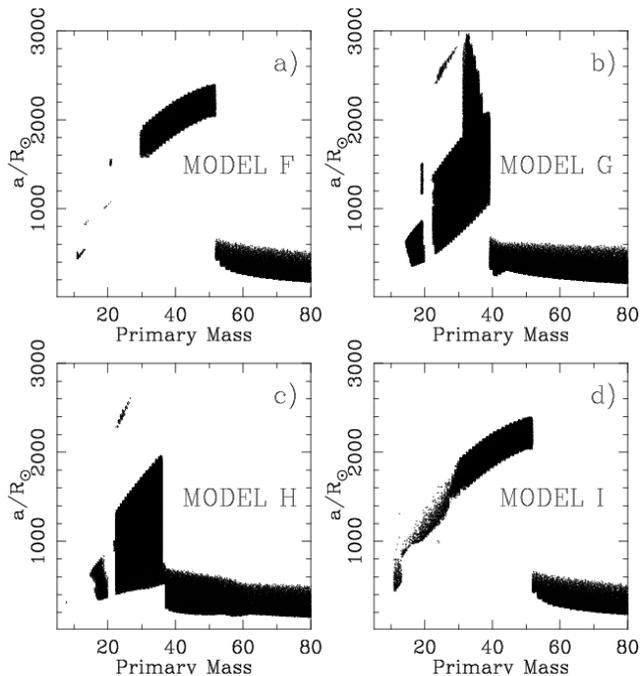}
  \caption{As Figure~\ref{f:fig7} but for Models~F, G, H and I. 
    \label{f:fig8}}
\end{figure}

Model H lowers the metallicity even further to a value of 0.0001 and we find that 
BH-MS-LMXB numbers have now increased to $6\,500$.  
The BH-MS-LMXB formation rate is now within a factor of 5 of its NS counterpart. 
Figure~8c) shows where the BH-LMXBs are created in Model H parameter space. 
The range in separation for BH-LMXBs, especially those with large 
primary masses, is increased as compared to both Model F and G. 
The BH-MS-LMXBs are also extended further down in primary mass to  
$23 M_\odot$ rather than $28 M_\odot$ for Model F. 
Within Figure~\ref{f:fig8}c) the BH-MS-LMXB systems are found within an area of
$23 M_\odot$ to $40 M_\odot$ in primary mass and $400 R_\odot$ to $1\,300 R_\odot$
in separation. 
We also find that there are a greater number of long lived LMXBs in Model~H 
than in Model~G. 
However, the evolution paths of binaries in Model~H differ less from what is 
found in Model~G compared to the differences observed between Models~F and G. 
An evolutionary example is given here to facilitate comparison between
Models F and H, and to some extent Model G.
Suppose a binary involves a primary mass of $M = 31.6 M_\odot$, secondary mass 
of $m = 0.95 M_\odot$ and separation of $a = 1243 R_\odot$.
Within Model H this binary produces a system, which after $6.8\,$Myr, 
goes into a CE phase with a CHeB primary of $M = 27 M_\odot$ and a separation 
of $1\,432 R_\odot$. 
The binary survives the CE with a separation of $5.7 R_\odot$. 
When the primary becomes an $8.6 M_\odot$ BH the separation is $8.0 R_\odot$. 
At a system time of $5\,380\,$Myr the $0.95 M_\odot$ secondary star evolves to 
fill its Roche-lobe and a BH-MS-LMXB system is born. 
After approximately $3\,000\,$Myr the $0.63 M_\odot$ secondary moves onto the 
HG ending its MS-LMXB phase. 
When this binary is evolved in Model~F the primary experiences more mass-loss while 
on the MS which means that the binary is wider than in Model~H when the primary 
reaches the end of the MS. 
However, the solar metallicity primary has a larger radius and fills its Roche-lobe while 
on the HG when the orbital separation is $1\,353 R_\odot$. 
A CE phase is initiated but the binary does not survive. 

Assuming that all stars within the Galaxy are born with the same metallicity is not 
the most realistic model as an age-metallicity relation clearly exists 
(e.g. Feltzing, Holmberg \& Hurley 2001). 
A simple extension of our models is to assume 
that the metallicity for all stars throughout the Galaxy is some 
percentage of the three different metallicity values we have used. 
If we take a mixture of 60\% of the stars having $Z = 0.02$, 30\% having $Z = 0.001$ 
and 10\% having $Z = 0.0001$ -- a somewhat arbitrary choice but in rough agreement 
with the metallicity distribution for stars in the Galaxy -- then with 
$\mu_{\rm nhe} = 0.5$ an estimated number of BH-MS-LMXBs is approximately 2759 
along with a birth rate of $1.194\times10^{-6}$. 
The corresponding NS-MS-LMXB formation rate is $7.61\times10^{-6}$ so the rates 
are within an order of magnitude which is pleasing. 
Of course we should correlate the stellar metallicity with the time of birth of each star, 
and consider a full range of metallicities, and this is something that could be done in 
the future. 

Another uncertain feature, which needs to be tested, is that of the 
velocity kick given at birth to compact objects.  
In the models presented so far we have chosen not to include 
kicks mainly to facilitate comparison with previous work 
(as discussed in Sec. 2.2). 
Also, due to the randomness inherent in modelling the 
velocity kicks neglecting this process makes 
comparisons between models more transparent. 
Giving a velocity kick to a star that has passed through a SN phase is, 
in effect, giving kinetic energy to that star and therefore kinetic 
energy to the system\footnote{In cases where the kick is directed 
in the opposite direction to the orbital motion of the star it 
is possible for the kinetic energy of the star to be reduced 
(e.g. Kalogera 1996).}. 
If the kinetic energy gained is greater than the gravitational potential 
energy of the system, then that system will become un-bound. 
Systems with low gravitational potential energy (i.e. larger separations) 
will be disrupted more easily than more strongly bound systems. 
The impact of allowing a velocity kick at the time of SN is investigated 
with Model~I. 
What we find is that the ratio of BH-LMXB systems with giant secondaries to those 
with MS star secondaries declines by a factor of two when going from Model~F to 
Model~I. 
Some of the BH-GB-LMXB systems which have been lost from Model I have 
instead produced BH-MS-LMXBs. 
One way in which this may occur is when a system that is too wide to form 
a BH-MS-LMXB normally has an eccentricity induced via the SN. 
Tidal forces then circularize the orbit and the separation decreases. 
In some cases the separation may be reduced sufficiently that the system can 
evolve to become a MS-LMXB. 
Alternatively one can imagine that in some systems the kick will widen the orbit to such an 
extent that mass-transfer does not occur, or occurs later in the evolution than expected 
for the same binary without a kick. 
Figure~8d) shows the parameter space in primary mass and orbital separation 
for Model~I and this can be compared to Figure~8a) for Model~F. 
The most noticeable change is the increase in Model~I of BH-LMXBs that 
evolved from NS-LMXBs and this is a direct consequence of the velocity kicks 
at birth for NSs increasing the likelihood of subsequent mass-transfer. 

\section{Discussion} 

In this research we have covered a substantial range of input parameters that 
affect the outcomes of binary evolution models. 
The standard scenario, Model A, gives both a poor ratio of birth rate between 
NS and BH MS-LMXBs and a low number of estimated BH-MS-LMXBs produced within the Galaxy. 
Both of these results differ to what observations suggest. 
Portegies Zwart, Verbunt \& Ergma (1997) find birthrates of 
$2.6\times10^{-6}$ for NS-MS-LMXBs and $9.6\times10^{-9}$ for BH-MS-LMXBs. 
So results for their standard model are similar to our Model~A. 
It is comforting that models using the same input parameters but different 
population synthesis codes produce results that are in agreement. 
However, both standard models give BH-LMXB formation rates much lower than 
what is indicated by observation. 
It is therefore a relief that our Models F, E and G give a reasonable match to the observations 
both for the rate of birth comparisons and actual numbers of systems. 
We must bear in mind that many areas of binary evolution remain uncertain and 
within population synthesis there are an uncomfortable number of parameters that are 
poorly constrained. 
A prime example is the modelling of the common-envelope phase. 
What we have been able to show is that no matter which of the algorithms for dealing 
with CE we use, or what value of the common-envelope efficiency parameter we 
adopt, it is not possible to make enough BH-LMXBs. 
So in effect our results are not affected by how the CE is modelled and uncertainty in 
this parameter becomes redundant. 
On the other hand we have shown that reducing the amount of mass-loss from helium 
stars does strongly affect the numbers of BH-MS-LMXB systems created. 
Simply reducing this by a factor of two from what had previously been used in the binary 
evolution algorithm -- a change that is allowed within the range indicated by 
observations of Wolf-Rayet stars (Pols \& Dewi 2002) 
-- we find that a healthy number of BH-MS-LMXBs are produced in Model~F. 
We therefore suggest that this represents a much better standard model for binary 
population synthesis calculations. 

The results that we have presented up to this point all assume that the age of the 
Galaxy is $15\,$Gyr which is a reasonable upper limit. 
To illustrate how the numbers react to a change in age we have repeated our 
analysis of Model~F but using $12\,$Gyr for the current age of the Galaxy. 
These results are given as Model~Fb in Table~2 and we see that there is a 
modest decrease in the birthrates and expected numbers. 
To investigate how a change in IMF slope affects the results we present a 
Model~Fc which takes a Salpeter (1955) IMF with a slope of $\alpha = 2.7$ 
as compared to $\alpha = 2.35$ for Model~F. 
In this case we find that the expected number of BH-MS-LMXBs in the 
Galaxy drops well below what is required to match the observations. 
Previous studies 
(e.g. Kalogera \& Webbink 1998; Pfahl, Rappaport \& Podsiadlowski 2003) 
have shown that the mass-ratio distribution assumed for the initial 
binary population can greatly affect the predictions regarding LMXB production.
Model~Fd in Table~2 is a repeat of Model~F but with the assumption that 
secondary star masses are not correlated with primary star masses 
-- in this case the secondary star masses are given by the 
Kroupa, Tout \& Gilmore (1993) IMF which is a much more accurate 
representation of the IMF for low-mass stars than the Salpeter (1955) IMF. 
We see that there is a modest increase in the BH-LMXB birthrate with a 
corresponding drop in the NS-LMXB rate and that the expected number of 
BH-LMXBs has roughly doubled. 
We note that repeating this change in the analysis for Model~A only increases 
the expected BH-LMXB number to 90 (from 17) which is still uncomfortably low. 

Another aspect that we need to check is that our choice of grid spacing has 
not affected the accuracy of our results. 
If a grid is too coarse, then potentially whole systems of interest may 
be excluded from the output and obviously we would like to avoid this. 
We have performed a Model~J which assumes the same input 
parameters as Model~F but uses a higher resolution grid where the spacing 
in both orbital separation and primary mass are halved. 
The results are given in Table~2 and are effectively the same as Model~F 
indicating that the results have converged and the size of the grid is not 
affecting the results. 

Pfahl, Rappaport \& Podsiadlowski (2003) have suggested 
that many LMXB systems are evolved  intermediate mass X-ray binaries (IMXBs). 
To allow for this we have performed one final model that extends the 
secondary mass range to include systems which evolve from secondaries 
initially more massive than $2 M_\odot$. 
This is Model K which deviates from Model F only by taking a 
secondary mass range of $0.1 M_\odot$ to $4 M_\odot$ with the number of grid 
points increased so that the stepsize of the two models is the same.  
A system with a secondary mass initially in excess of $2 M_\odot$ can contribute as an 
LMXB if the secondary mass drops below $2 M_\odot$ during mass-transfer. 
The results for Model~K are given in Table~2. 
The formation rates calculated from this model have increased 
compared to Model F with the NS-MS-LMXB rate of production being 
$2.2 \times 10^{-5}$ and the BH-MS-LMXB rate of production at $9.2 \times 10^{-7}$. 
However, the ratio of NS to BH LMXB birthrates has decreased compared to Model~F. 
The number of BH-MS-LMXB systems has not changed significantly and is now $1910$.
Within Model K there is a greater range of LMXB production from very high mass primary 
star systems ($M > 50 M_\odot$) -- these now cover separations between  
$200 < a < 1400 R_\odot$.
Also, the range of initial separations for systems with a primary mass of $30 - 50  M_\odot$ 
has widened to $800 - 2600 R_\odot$. 
These are significant parameter space increases when compared to Model F.

Figure~9 compares the distributions of parameter space for 
Models~F and K at the initiation of the BH-LMXB phase. 
The greater numbers of BH-LMXB systems produced by Model K as compared 
with Model F can be easily seen. 
The BH-MS-LMXB region for both models in Figure~9 is at low periods 
between 0.1 and 1 day and covers 0.2 to $2 M_\odot$ in secondary mass. 
Thus the models suggest that BH-LMXBs observed to have periods in 
excess of $1\,$d are likely to have 
sub-giant or giant donors as opposed to MS star companions. 
We also note that unlike Kalogera \& Webbink (1998) we do not 
need velocity kicks to produce systems with orbital periods 
less than $1\,$d. 
Presumably this is a result of reducing the mass-loss rate 
from helium stars as well as differences in the modelling 
of aspects such as common-envelope evolution and tides. 

Model K  does produce an increase in the numbers of BH-MS-LMXB systems but not 
by as much as may have been envisaged (e.g. Pfahl, Rappaport \& Podsiadlowski 2003). 
This arises because the BH-MS-LMXB phase life-times of the 
additional systems that evolve from IMXBs are relatively short lived.
For example, a 
system with an initial primary star mass of $46 M_\odot$, a secondary 
star mass of $2.4 M_\odot$ and a binary separation of $1910 R_\odot$ 
becomes a MS-IMXB at a system time of 417 Myr, where the secondary star 
mass has increased slightly to $2.45 M_\odot$.
With the secondary star losing mass to the BH primary its mass eventually
falls below the low-mass boundary of $2.0 M_\odot$ at $660\,$Myr.
This initiates a BH-MS-LMXB phase which ends when the secondary evolves 
off the MS and results in a phase life-time of approximately $80\,$Myr. 
The MS-LMXB phase is only a quarter of the total X-ray binary life-time.
So systems which evolve from IMXBs have higher initial secondary masses 
than those that start mass-transfer as LMXBs and thus the MS lifetimes are 
shorter and also time is taken up by the secondary reducing its mass below 
the $2 M_\odot$ LMXB cutoff. 
Previously the IMXB to LMXB formation channel has been utilised to correct for the 
under-production of standard LMXBs but here this is not necessary as with Model~F 
we already have a healthy MS-LMXB population. 
The extension to include descendants of IMXBs still gives us a BH-MS-LMXB number 
that is a good match to the observed range. 

\begin{figure*}
  \includegraphics[width=168mm]{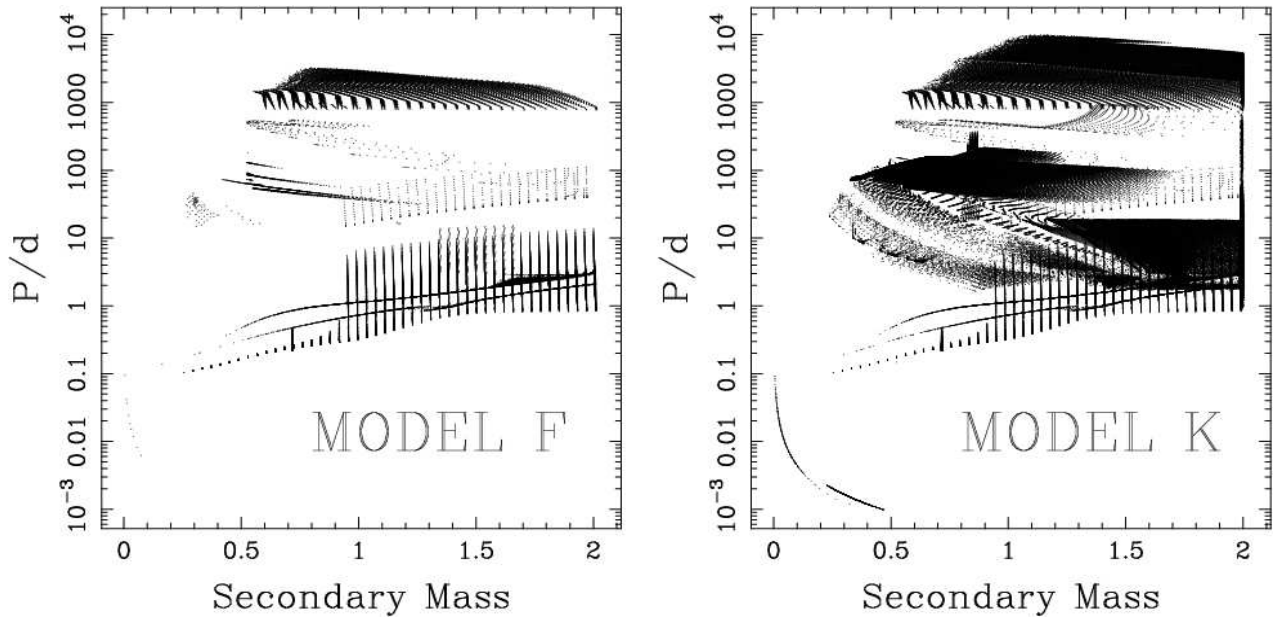}
  \caption{Parameter space of secondary mass and orbital period at the 
           start of the BH-LMXB phase for systems formed in 
           Model~F (left panel) and in Model~K (right panel). 
           The lined appearance of the points is due to the grid approach of our method.
    \label{f:fig9}}
\end{figure*}

In this work we have compared our results to the estimated number of 
BH-MS-LMXB systems in the Galaxy. 
This number is based on observations and attempts to account for selection effects and 
the possible transient nature of the sources. 
An alternative is to approach the problem from the opposite side and force the 
population synthesis results to reflect actual surveys 
-- including differentiation between transient and persistent sources. 
This method would potentially reduce the uncertainties involved in the comparison 
and help to better constrain the evolutionary parameters. 
Particularly as most BH-LMXB systems are found to be transient systems 
as opposed to NS-LMXB systems which are generally persistant sources 
(van Paradijs 1996; King, Kolb \& Szuszkiewicz 1997; Chen, Shrader \& Livio 1997).
Transient sources have typical outburst times of 100 days while the quiescent
phase is of order several years (Truss et al. 2002) and obviously this affects 
whether or not a system is observable. 
At present there seems to be two methods to model such transient
features in a rapid binary evolution code.
These rely on comparing the mass transfer rate via RLOF to a
critical mass loss rate (King, Kolb \& Szuszkiewicz 1997) or to compare
the luminosity of the accretion disk to a critical luminosity (van Paradijs 1996).
This is an approach we plan to take in future work but only after producing 
detailed models of accretion in LMXB systems in order to get a better handle 
on transient timescales and how these vary with system parameters. 
The possibility of irradiation affecting transient behaviour will also need to 
be included (Pfahl, Rappaport \& Podsiadlowski 2003). 
A related extension is to follow the spatial trajectories of the LMXBs within the 
Galactic potential while also following the binary evolution, 
as suggested previously by Pfahl, Rappaport \& Podsiadlowski (2003). 
The resulting population can then be surveyed taking into account real 
selection effects. 
This will also enable more reliable comparison between synthesized and 
observed populations. 

Finally we mention some alternative theories for close binary formation 
related to the formation of LMXB systems. 
These are the triple system scenario (Eggleton \& Verbunt 1986) and the 
tidal capture scenario (Fabian, Pringle \& Rees 1975). 
The triple scenario involves two massive stars in a short-period binary which 
comprises the inner binary of a triple system with a low-mass dwarf star. 
The most massive star collapses to a BH but owing to its companion being 
both close and massive it is likely that the binary remains intact. 
Eventually the binary components merge to form a BH surrounded by an 
envelope which expands on a thermal timescale. 
This envelope engulfs the third star to setup a common-envelope 
situation in which the low-mass dwarf spirals-in towards the BH. 
Survival of this CE phase leaves a close binary which may then form an LMXB 
if the low-mass star subsequently fills its Roche-lobe. 
The second alternative theory considers a close encounter between a compact 
object that approaches within a few $R_\odot$ of a non-compact object. 
During this encounter the gravitational energy between the two stars is absorbed 
by oscillations produced in the non-compact object. 
If enough energy is absorbed in exiting these oscillations then the two stars 
can become bound, and as shown by Fabian, Pringle \& Rees (1975) the system created 
is a close binary. 
The evolution of the non-compact object may then lead to RLOF and an LMXB phase 
(if the non-compact object is $< 2 M_\odot$). 
This type of close binary formation is only possible in dense stellar regions, 
as found in the cores of globular clusters. As Eggleton \& Verbunt (1986) explain, 
the number of close encounters is too small, in the galactic disk or bulge, to 
account for the numbers of LMXB systems seen.
Another possibility that exists in the dense stellar environment of a star 
cluster is the capture of neutron stars into close binaries through exhange 
interactions (Hut, Murphy \& Verbunt 1991). 

\section{Conclusions}

To model the Galactic population of low-mass X-ray binaries we 
have performed population synthesis using a binary evolution algorithm that 
follows tidal circularization and synchronisation of binary orbits. 
Our favoured model (Model~F) estimates that the Galaxy harbours about $1900$ LMXBs 
containing a BH and a MS star companion. 
The same model predicts the BH-MS-LMXB birthrate to be  
$4.4 \times 10^{-7} \rm{yr}^{-1}$ and the NS-MS-LMXB birthrate to be 
$6.5 \times 10^{-6} \rm{yr}^{-1}$. 
This is in good agreement with the inference from observational surveys 
that the number of BH-MS-LMXBs residing in the Galaxy is of the order 
of thousands and that the birthrates of LMXBs with NS and BH components 
should be comparable. 
However, we emphasise that these are not robust constraints owing 
to the uncertainties involved. 

To achieve the results of our favoured model we made one simple change to the 
set of input parameters 
to the BSE algorithm that had previously been assumed as standard 
(Hurley, Tout \& Pols 2002). 
Namely we altered the mass-loss rate from helium stars to reflect the improvement 
suggested by Hamann \& Koesterke (1998) over the original prescription of 
Hamann, Koesterke \& Wessolowski (1995) which had until now been adopted 
in BSE. 
This amounted to reducing the helium star mass-loss strength by a factor of two. 
Without this reduction our standard model (Model~A) predicted of the order of 10  
BH-MS-LMXBs in the Galaxy -- failing miserably to match the order of magnitude 
estimate suggested by observations. 
This is in line with the findings of previous population synthesis calculations 
(Portegies Zwart, Verbunt \& Ergma 1997; Podsiadlowski, Rappaport \& Han 2003) 
using standard assumptions regarding binary evolution. 

We have also investigated the importance of a number of other parameters involved 
in binary evolution in determining the population synthesis outcome. 
Varying the common-envelope efficiency parameter, even to excessively large values, 
does not lead to anywhere near enough BH-MS-LMXBs (Models~C and D). 
Changing the way in which the CE is modelled (Model~B) does not improve on the standard 
model result either. 
It is actually reassuring that modelling of this ad-hoc process and the choice of 
the very uncertain efficiency parameter does not have a strong bearing on the results 
-- although obviously a very low efficiency will be detrimental to LMXB production 
irrespective of the choices made for other parameters. 
Altering the mass-ratio distribution for the initial binary population 
(from correlated to uncorrelated masses) does not boost the BH-MS-LMXB 
numbers of the standard model past 100. 
As expected we do find that lowering the maximum neutron star mass from $3.0$ 
to $2.0 M_\odot$ (Model~E) boosts BH-MS-LMXB numbers and gives a good match to the 
expectations derived from LMXB surveys. 
However, this change to the model is not as preferable as reducing the helium star wind 
as for the latter there is observational evidence to justify the change. 
We find that including LMXBs that evolve from IMXBs is not necessary to explain the 
observed number of BH-LMXBs but this is a natural evolution  
sequence and cannot be neglected in these and future calculations. 

For the majority of our models we have neglected to impart a kick 
velocity at birth for the neutron stars and black holes. 
This was to facilitate comparision with previous studies that took 
this approach. 
It also aids in directly quantifying how varying evolution 
parameters from model to model affects LMXB production.  
However, for neutron stars at least, there is compelling 
observational evidence for a velocity kick at birth and this 
needs to be accounted for in LMXB population synthesis 
(Kalogera \& Webbink 1998). 
Taking our favoured model and including velocity kicks (Model~I) 
increased the production of BH-LMXBs but did not harm the 
compatibility of the results. 
In fact it also gave a NS-LMXB birthrate that compares well 
with the Galactic birthrate of binary millisecond pulsars. 
We find that we do not require velocity kicks in order 
to produce short-period BH-LMXBs. 

We suggest that reducing the Hamann, Koesterke \& Wessolowski (1995) 
helium-star mass-loss prescription by a factor of two should be adopted as 
standard in future population synthesis calculations. 
Alternatively the weaker rate suggested by Nugis \& Lamers (2000) should 
be used. 
We also find that varying the initial metallicity of the population can strongly 
affect the LMXB results and therefore including the Galactic age-metallicity 
relation in future models would be desirable. 
Another extension of this work is to model the spatial distribution of LMXBs 
in the Galaxy. 
This will allow us to conduct synthetic surveys of the population and compare 
with raw observed numbers rather than using numbers that have been 
manipulated to correct for observational uncertainties. 
In this way we will be able to place stronger constraints on uncertain 
parameters in the binary modelling process and the results 
would also be advantageous to observers for understanding existing data 
and shaping future surveys.

\section*{Acknowledgments}

We thank Onno Pols, James Murray and Matthew Bailes for helpful discussions 
during this work. 
We are also grateful to the two referees for comments that 
improved the focus of the work. 
PDK thanks the Centre for Astrophysics and Supercomputing, 
Swinburne University of Technology for a scholarship.
JRH thanks the Australian Research Council for a Fellowship.

\begin{table*}
 \centering
 \scriptsize
 \begin{minipage}{140mm}
  \caption{Charactersitics of the main set of models used in this work. 
                 The first row gives an identifying letter for each model. 
                 In the second row we indicate the value of the common-envelope efficiency 
                 parameter and this is followed by the choice of common-envelope scheme. 
                 The fourth row shows the metallicity used. 
                 Next is the helium wind parameter and the maximum mass assumed for 
                 neutron stars. 
                 Finally we indicate whether or not kicks were allowed at birth for neutron stars. 
  \label{t:table1}}
  \begin{tabular}{crrrrrrrrrrrrrr}
  \hline
 Model & A & B & C & D & E & F & G & H & I \\
\hline 
 $\alpha_{\rm\SSS CE}$ & 3 & 3 & 10 & 1 & 3 & 3 & 3 & 3 & 3 \\
 CE scheme & BSE & PRH & BSE & BSE & BSE & BSE & BSE & BSE & BSE \\
 Z & 0.02 & 0.02 & 0.02 & 0.02 & 0.02 & 0.02 & 0.001 & 0.0001 & 0.02 \\
 $\mu_{\rm nhe}$ & 1 & 1 & 1 & 1 & 1 & 0.5 & 0.5 & 0.5 & 0.5 \\
 $M_{max,NS}$ & 3 & 3 & 3 & 3 & 2 & 3 & 3 & 3 & 3 \\
 Kick & No & No & No & No & No & No & No & No & Yes \\
\hline
\end{tabular}
\end{minipage}
\end{table*}

\begin{table*}
 \centering
 \begin{minipage}{140mm}
  \caption{Formation rates of NS and BH MS-LMXBs and the predicted numbers of 
                  BH-MS-LMXBs in the Galaxy for a variety of models. 
                  The age of the Galaxy is assumed to be $15\,$Gyr in all cases except 
                  for Model~Fb which assumes an age of $12\,$Gyr and is identical to 
                  Model~F in all other respects. 
                  Model~Fc varies from Model~F only by assuming that the slope of the Salpeter 
                  IMF is 2.7 as opposed to the value of 2.35 used otherwise. 
                  Model~Fd varies from Model~F by assuming that the secondary mass is 
                  drawn independently from the Kroupa, Tout \& Gilmore (1993) IMF. 
                  Model~J is the same as Model~F except that a finer grid is used and Model~K 
                  allows for LMXBs to be descendants of IMXBs. 
  \label{t:table2}}
  \begin{tabular}{crrr}
  \hline
 Model & \multicolumn{2}{c}{MS-LMXB rate/yr$^{-1}$} & Number \\
 &  \multicolumn{1}{c}{NS} &  \multicolumn{1}{c}{BH} & \\
\hline 
 A & $6.4 \times 10^{-6}$ & $3.1 \times 10^{-9}$ & $17$ \\
 B & $9.5 \times 10^{-6}$ & $3.1 \times 10^{-9}$  & $17$ \\
 C & $4.6 \times 10^{-6}$ & $4.3 \times 10^{-8}$  & $220$ \\
 D & $1.5 \times 10^{-6}$ & $0.0 \times 10^{-0}$  & $0$ \\
 E & $6.3 \times 10^{-6}$ & $4.6 \times 10^{-7}$  & $1470$ \\
 F & $6.5 \times 10^{-6}$ & $4.4 \times 10^{-7}$  & $1870$ \\
 Fb & $5.5 \times 10^{-6}$ & $4.1 \times 10^{-7}$ & $1340$ \\
 Fc & $1.6 \times 10^{-6}$ & $6.6 \times 10^{-8}$ & $280$ \\
 Fd & $2.8 \times 10^{-6}$ & $5.3 \times 10^{-7}$ & $4060$ \\
 G & $5.7 \times 10^{-6}$ & $1.7 \times 10^{-6}$  & $3290$ \\
 H & $2.0 \times 10^{-5}$ & $4.2 \times 10^{-6}$  & $6500$ \\
 I & $2.2 \times 10^{-5}$ & $1.0 \times 10^{-6}$  & $6130$ \\
 J & $6.2 \times 10^{-6}$ & $4.4 \times 10^{-7}$  & $1860$ \\
 K & $2.2 \times 10^{-5}$ & $9.2 \times 10^{-7}$ & $1910$ \\
\hline
\end{tabular}
\end{minipage}
\end{table*}

\bsp

\label{lastpage}

\end{document}